\documentclass[sigconf,nonacm]{acmart}

\usepackage{booktabs}
\usepackage{amsmath}
\usepackage{graphicx}
\usepackage{placeins}
\graphicspath{{figures/}}
\newcommand{\BaseVideos}{5,672}
\newcommand{\BaseCategories}{38}
\newcommand{\BaseRandomRows}{893,793}
\newcommand{\BaseRandomUsers}{26,998}
\newcommand{\BaseStandardRows}{219,707}
\newcommand{\BaseStandardUsers}{24,483}
\newcommand{\EligibleStandardUsers}{12,477}

\newcommand{\EligibleStandardUserShare}{50.96}
\newcommand{\PairedBaseUserShare}{50.94}
\newcommand{\PairedBaseVideoShare}{96.72}
\newcommand{\PairedUsers}{12,472}
\newcommand{\PairedVideos}{5,486}
\newcommand{\PairedRandomRows}{467,919}
\newcommand{\PairedStandardRows}{180,645}
\newcommand{\NaiveAbs}{0.1935}
\newcommand{\NaiveCentered}{0.0491}
\newcommand{\NaiveMaximum}{0.3037}
\newcommand{\NaiveRank}{0.7833}
\newcommand{\NaiveTopTen}{0.70}
\newcommand{\DemeanAbs}{0.0387}
\newcommand{\DemeanCentered}{0.0387}
\newcommand{\DemeanMaximum}{0.0996}
\newcommand{\DemeanRank}{0.7888}
\newcommand{\DemeanTopTen}{0.80}
\newcommand{\AbsoluteReduction}{80.0}
\newcommand{\CenteredReduction}{21.0}
\newcommand{\RandomGrandMean}{0.133071}
\newcommand{\StandardGrandMean}{0.336268}

\newcommand{\AbsDifference}{-0.1548}
\newcommand{\AbsDifferenceCI}{[-0.1587, -0.1499]}
\newcommand{\CenteredDifference}{-0.0103}
\newcommand{\CenteredDifferenceCI}{[-0.0131, -0.0072]}
\newcommand{\RankDifference}{0.0055}
\newcommand{\RankDifferenceCI}{[-0.0217, 0.0412]}
\newcommand{\MainBootstrapDraws}{1,000}
\newcommand{\ResponseShapley}{0.1409}
\newcommand{\ItemMixShapley}{0.0526}
\newcommand{\ResponseShapleyShare}{72.8}
\newcommand{\ItemMixShapleyShare}{27.2}
\newcommand{\CenteredResidualShapley}{0.0239}
\newcommand{\CenteredItemMixShapley}{0.0252}
\newcommand{\CenteredResidualShare}{48.6}
\newcommand{\CenteredItemMixShare}{51.4}
\newcommand{\MainShapleyBootstrapDraws}{1,000}
\newcommand{\AccountingIdentityError}{$5.6\times10^{-17}$}
\newcommand{\WithinVideoMeanAbs}{0.1445}
\newcommand{\ItemMixMeanAbs}{0.0561}
\newcommand{\WithinVideoDominatesCount}{36}
\newcommand{\AccountingCategoryCount}{38}

\newcommand{\AbsoluteResidualSharePctCI}{[71.6, 74.0]}
\newcommand{\AbsoluteItemMixSharePctCI}{[26.0, 28.4]}
\newcommand{\CenteredResidualSharePctCI}{[44.9, 53.0]}
\newcommand{\CenteredItemMixSharePctCI}{[47.0, 55.1]}
\newcommand{\FullSupportAbs}{0.2000}
\newcommand{\FullSupportCentered}{0.0512}
\newcommand{\FullSupportRank}{0.7527}
\newcommand{\CommonRandomRowShare}{75.97}
\newcommand{\CommonStandardRowShare}{99.98}

\newcommand{\FullRandomVideos}{7,274}
\newcommand{\FullStandardVideos}{5,494}
\newcommand{\CommonSupportVideos}{5,486}
\newcommand{\ThresholdUsersOne}{24,457}
\newcommand{\ThresholdUsersTwo}{21,325}
\newcommand{\ThresholdUsersThree}{18,261}
\newcommand{\ThresholdUsersFive}{12,472}
\newcommand{\ThresholdUsersTen}{3,585}
\newcommand{\ThresholdNaiveAbsOne}{0.1997}
\newcommand{\ThresholdNaiveAbsTwo}{0.1993}
\newcommand{\ThresholdNaiveAbsThree}{0.1975}
\newcommand{\ThresholdNaiveAbsFive}{0.1935}
\newcommand{\ThresholdNaiveAbsTen}{0.1818}
\newcommand{\ThresholdNaiveCenteredOne}{0.0523}
\newcommand{\ThresholdNaiveCenteredTwo}{0.0519}
\newcommand{\ThresholdNaiveCenteredThree}{0.0509}
\newcommand{\ThresholdNaiveCenteredFive}{0.0491}
\newcommand{\ThresholdNaiveCenteredTen}{0.0475}
\newcommand{\ThresholdNaiveRankOne}{0.7472}
\newcommand{\ThresholdNaiveRankTwo}{0.7582}
\newcommand{\ThresholdNaiveRankThree}{0.7639}
\newcommand{\ThresholdNaiveRankFive}{0.7833}
\newcommand{\ThresholdNaiveRankTen}{0.8028}
\newcommand{\ThresholdDemeanAbsOne}{0.0380}
\newcommand{\ThresholdDemeanAbsTwo}{0.0388}
\newcommand{\ThresholdDemeanAbsThree}{0.0388}
\newcommand{\ThresholdDemeanAbsFive}{0.0387}
\newcommand{\ThresholdDemeanAbsTen}{0.0400}
\newcommand{\ThresholdNaiveAbsRange}{0.1818--0.1997}
\newcommand{\ThresholdDemeanAbsRange}{0.0380--0.0400}
\newcommand{\RowDemeanAbs}{0.0368}
\newcommand{\RowDemeanCentered}{0.0368}
\newcommand{\RowDemeanRank}{0.8094}
\newcommand{\WeightedNaiveAbs}{0.2036}
\newcommand{\WeightedNaiveCentered}{0.0487}
\newcommand{\WeightedDemeanAbs}{0.0311}
\newcommand{\WeightedDemeanCentered}{0.0310}
\newcommand{\TemporalUsers}{8,983}
\newcommand{\TemporalVideos}{3,680}
\newcommand{\TemporalBootstrapDraws}{1,000}
\newcommand{\EarlyNaiveAbs}{0.1784}
\newcommand{\LateNaiveAbs}{0.1732}
\newcommand{\EarlyNaiveCentered}{0.0511}
\newcommand{\LateNaiveCentered}{0.0478}
\newcommand{\EarlyNaiveRank}{0.7223}
\newcommand{\LateNaiveRank}{0.8033}
\newcommand{\TemporalAbsChange}{-0.0052}
\newcommand{\TemporalAbsChangeCI}{[-0.0130, 0.0023]}
\newcommand{\TemporalCenteredChange}{-0.0033}
\newcommand{\TemporalCenteredChangeCI}{[-0.0101, 0.0040]}
\newcommand{\TemporalRankChange}{0.0810}
\newcommand{\TemporalRankChangeCI}{[-0.0247, 0.1315]}
\newcommand{\TemporalGapRank}{0.8146}
\newcommand{\TemporalGapRankCI}{[0.6555, 0.8718]}
\newcommand{\ClickNaiveAbs}{0.2915}
\newcommand{\ClickNaiveCentered}{0.0446}
\newcommand{\ClickNaiveRank}{0.4413}
\newcommand{\ClickDemeanAbs}{0.0339}
\newcommand{\ClickDemeanCentered}{0.0323}
\newcommand{\ClickDemeanRank}{0.3866}
\newcommand{\ClickNaiveMaximum}{0.4540}
\newcommand{\ClickNaiveTopTen}{0.60}
\newcommand{\ClickDemeanMaximum}{0.1001}
\newcommand{\ClickDemeanTopTen}{0.50}
\newcommand{\LongNaiveAbs}{0.2389}
\newcommand{\LongNaiveCentered}{0.0431}
\newcommand{\LongNaiveRank}{0.4711}
\newcommand{\LongNaiveMaximum}{0.3767}
\newcommand{\LongNaiveTopTen}{0.50}
\newcommand{\LongDemeanAbs}{0.0278}
\newcommand{\LongDemeanCentered}{0.0255}
\newcommand{\LongDemeanRank}{0.4660}
\newcommand{\LongDemeanMaximum}{0.0856}
\newcommand{\LongDemeanTopTen}{0.40}
\newcommand{\UnclippedNaiveAbs}{0.2276}
\newcommand{\UnclippedNaiveCentered}{0.0650}
\newcommand{\UnclippedNaiveRank}{0.8103}
\newcommand{\UnclippedDemeanAbs}{0.0519}
\newcommand{\UnclippedDemeanCentered}{0.0519}
\newcommand{\UnclippedDemeanRank}{0.7890}
\newcommand{\ConfidenceUsers}{12,443}
\newcommand{\ConfidenceVideos}{5,468}
\newcommand{\ConfidenceNaiveAbs}{0.1934}
\newcommand{\ConfidenceNaiveCentered}{0.0490}
\newcommand{\ConfidenceNaiveRank}{0.7746}
\newcommand{\ConfidenceDemeanAbs}{0.0399}
\newcommand{\ConfidenceDemeanCentered}{0.0399}
\newcommand{\ConfidenceDemeanRank}{0.7650}
\newcommand{\LedgerRandomOnlyVideos}{1,788}
\newcommand{\LedgerStandardOnlyVideos}{8}
\newcommand{\LedgerFullWeightedMeanGap}{0.2102}
\newcommand{\LedgerWeightedRMSE}{0.2182}
\newcommand{\LedgerBootstrapDraws}{1,000}
\newcommand{\LedgerWithinTotalShare}{72.41}
\newcommand{\LedgerMixTotalShare}{23.90}
\newcommand{\LedgerRandomSupportTotalShare}{3.65}
\newcommand{\LedgerStandardSupportTotalShare}{0.035}
\newcommand{\LedgerWithinShapeShare}{62.68}
\newcommand{\LedgerMixShapeShare}{31.39}
\newcommand{\LedgerRandomSupportShapeShare}{5.78}
\newcommand{\LedgerStandardSupportShapeShare}{0.15}
\newcommand{\LedgerWithinTotalShareCI}{[71.54, 73.33]}
\newcommand{\LedgerMixTotalShareCI}{[23.05, 24.75]}
\newcommand{\ProspectiveHistoryEligibleUsers}{20,178}
\newcommand{\ProspectiveFrozenUsers}{14,912}
\newcommand{\ProspectiveTestUsers}{12,342}
\newcommand{\ProspectiveFrozenVideos}{4,852}
\newcommand{\ProspectiveTestRandomVideos}{4,752}
\newcommand{\ProspectiveTestStandardVideos}{3,839}
\newcommand{\ProspectiveFrozenRetention}{82.8}
\newcommand{\ProspectiveHistoryRetention}{61.2}
\newcommand{\ProspectiveBootstrapDraws}{1,000}
\newcommand{\ProspectiveNaiveAbs}{0.1853}
\newcommand{\ProspectiveScalarAbs}{0.0531}
\newcommand{\ProspectiveScalarCentered}{0.0515}
\newcommand{\ProspectiveCategoryAbs}{0.0251}
\newcommand{\ProspectiveCategoryCentered}{0.0261}
\newcommand{\ProspectiveCategoryRank}{0.8545}
\newcommand{\ProspectiveCategoryTopTen}{0.90}
\newcommand{\ProspectiveRWAbs}{0.0675}
\newcommand{\ProspectiveRWCentered}{0.0409}
\newcommand{\ProspectiveCategoryMinusScalarAbs}{-0.0280}
\newcommand{\ProspectiveCategoryMinusScalarAbsCI}{[-0.0296, -0.0172]}
\newcommand{\ProspectiveCategoryMinusScalarCentered}{-0.0255}
\newcommand{\ProspectiveCategoryMinusScalarCenteredCI}{[-0.0288, -0.0159]}
\newcommand{\ProspectiveGapRank}{0.8678}
\newcommand{\DecisionBootstrapDraws}{1,000}
\newcommand{\DecisionFiveNaiveRegret}{0.0093}
\newcommand{\DecisionFiveCERegret}{0.0146}
\newcommand{\DecisionTenNaiveRegret}{0.0162}
\newcommand{\DecisionTenCERegret}{0.0101}
\newcommand{\DecisionTenRWRegret}{0.0101}
\newcommand{\DecisionTenNaiveRegretCI}{[0.0100, 0.0180]}
\newcommand{\DecisionTenRWRegretCI}{[0.0072, 0.0114]}
\newcommand{\DecisionTenRWChange}{-0.0062}
\newcommand{\DecisionTenRWChangeCI}{[-0.0084, -0.0003]}
\newcommand{\DecisionTenNaiveOverlap}{0.70}
\newcommand{\DecisionTenRWOverlap}{0.80}
\newcommand{\DecisionFiveRWChangeCI}{[-0.0217, 0.0154]}
\newcommand{\DecisionTwentyRWChangeCI}{[-0.0061, 0.0021]}
\newcommand{\DecisionTenRWFamilywiseCI}{[-0.0136, 0.0013]}
\newcommand{\DecisionFamilywiseCritical}{3.492}
\newcommand{\DecisionNaiveCenteredBandExclusions}{24}
\newcommand{\DecisionCECenteredBandExclusions}{27}
\newcommand{\DecisionRWCenteredBandExclusions}{24}
\newcommand{\BenchmarkReplications}{100}
\newcommand{\BenchmarkRowsPerReplication}{60,000}
\newcommand{\SplitReferenceSelectionRows}{233,793}
\newcommand{\SplitReferenceEvaluationRows}{234,126}

\newcommand{\SplitReferenceTenRWChange}{-0.0061}
\newcommand{\SplitReferenceTenRWChangeCI}{[-0.00840, 0.00003]}
\newcommand{\SplitReferenceTenRWFamilywiseCI}{[-0.0136, 0.0014]}
\newcommand{\SplitReferencePointDirectionsMaintained}{6}

\newcommand{\DependenceDateOmissions}{17}
\newcommand{\DependenceDateCenteredChangeRange}{[-0.0116, -0.0101]}
\newcommand{\DependenceDateCenteredReductionRange}{[20.8\%, 23.5\%]}
\newcommand{\DependenceVideoDraws}{500}
\newcommand{\DependenceVideoCenteredEnvelope}{[-0.0233, -0.0047]}
\newcommand{\DependenceVideoCenteredFavorableShare}{99.6}
\newcommand{\DependenceVideoRankFavorableShare}{53.2}
\newcommand{\DependenceVideoTopTenFavorableShare}{61.2}
\newcommand{\BenchmarkTargetNaiveCentered}{0.0146}
\newcommand{\BenchmarkUserNaiveCentered}{0.0297}
\newcommand{\BenchmarkUserRWCentered}{0.0163}
\newcommand{\BenchmarkInteractionNaiveCentered}{0.0247}
\newcommand{\BenchmarkInteractionRWCentered}{0.0151}
\newcommand{\BenchmarkItemNaiveCentered}{0.0218}
\newcommand{\BenchmarkItemRWCentered}{0.0214}
\newcommand{\BenchmarkWeakRWCentered}{0.0207}
\newcommand{\BenchmarkWeakOracleCentered}{0.0522}

\newcommand{\BenchmarkWeakMinCategoryESS}{8.7}
\newcommand{\BenchmarkPositiveNaiveCentered}{0.0311}
\newcommand{\BenchmarkPositiveRWCentered}{0.0159}
\newcommand{\BenchmarkOracleAtSixty}{0.0540}
\newcommand{\BenchmarkOracleAtThreeHundred}{0.0256}

\title{When Category Means Move: A Paired-Support Audit of
Recommendation Logs with Random Insertions}

\author{Yu Zhang}
\orcid{0009-0003-5368-1356}
\affiliation{\institution{AMap Alibaba Group}\city{Beijing}\country{China}}
\email{yuanyu.zy@alibaba-inc.com}

\author{Zhihan Wang}
\orcid{0009-0000-8880-7172}
\affiliation{\institution{AMap Alibaba Group}\city{Beijing}\country{China}}
\email{yingchen.wzh@alibaba-inc.com}

\author{Yuhang Lu}
\orcid{0009-0006-6883-2227}
\affiliation{\institution{AMap Alibaba Group}\city{Beijing}\country{China}}
\email{lyh203611@alibaba-inc.com}

\author{Shuai Li}
\orcid{0009-0002-0986-2881}
\affiliation{\institution{AMap Alibaba Group}\city{Beijing}\country{China}}
\email{lion.lis@alibaba-inc.com}

\hypersetup{pdfauthor={Yu Zhang, Zhihan Wang, Yuhang Lu, Shuai Li}}

\begin{document}
\raggedbottom
\pagestyle{plain}

\begin{abstract}
Content-category dashboards are built from production recommendation logs,
although the serving policy selects the users, items, and contexts behind each
mean.  Using KuaiRand random-insertion impressions, we audit such summaries
after fixing the same \PairedUsers{} users and \PairedVideos{} videos in both
logs.  Standard means have absolute/centered mean absolute error (MAE)
\NaiveAbs{}/\NaiveCentered{} and rank correlation \NaiveRank{}.  User demeaning
plus a random grand-mean anchor reduces absolute MAE by \AbsoluteReduction\%,
but centered MAE by only \CenteredReduction\% and does not reliably improve
rank.  A
complete-support ledger allocates weighted squared error mainly to within-video
residual (\LedgerWithinTotalShare\%) and video mix (\LedgerMixTotalShare\%),
with \LedgerRandomSupportTotalShare\% from random support.  In a frozen later
test, an early category calibration reaches MAE \ProspectiveCategoryAbs{}
versus \ProspectiveScalarAbs{} for scalar calibration; their direct difference
is \ProspectiveCategoryMinusScalarAbs{} \ProspectiveCategoryMinusScalarAbsCI{}.
Decision evidence is cutoff-specific: the nominal top-10 regret change for
row-weighted (RW) demeaning versus naive is \DecisionTenRWChange{}
\DecisionTenRWChangeCI{}, while top-5 and
top-20 intervals include zero; the six-contrast family-wise top-10 interval is
\DecisionTenRWFamilywiseCI{} and includes zero.  An outcome-free disjoint
randomized-reference split leaves all six nominal contrasts covering zero as
well.  A known-truth stress test shows incomplete
improvement under a mean-additive positive control and exposes failure under
item mix and weak overlap.  All decompositions are descriptive and every result
is conditional on its overlap cohort.  KuaiRand intervals quantify user-cluster
variation conditional on the realized catalog and calendar; we claim neither
intrinsic category effects nor a universally valid correction.
\end{abstract}


\maketitle

\section{Introduction}

Recommendation dashboards often rank content categories by mean engagement
and use that ranking for sourcing, promotion, and inventory decisions.  The
means are simple; their interpretation is not.  A serving policy changes the
audience, frequency, context, and within-category item mix represented in a
log.  A dashboard can therefore move without an underlying change in the
content category itself.  This is a measurement consequence of policy-shaped
feedback, related to but distinct from learning and evaluating a recommender
under missing-not-at-random interactions~\cite{chaney2018algorithmic,castells2022offline}.

We ask a measurement question: \emph{how far are production-log category
summaries from a matched random-insertion reference, and which parts of the
gap can a simple user adjustment change?}  A single error number is
insufficient.  A bucket-wide level shift can produce a large absolute gap even
when rankings are similar, whereas an operational top-$k$ decision depends on
category shape.  We therefore report absolute MAE, centered-shape MAE, maximum
error, rank correlation, and top-10 overlap separately.

The main design uses KuaiRand~\cite{gao2022kuairand}.  We align random-insertion
and standard-serving impressions on dates, interface, official category
mapping, users, and videos.  Restricting to common videos changes the estimand:
it removes support entry and exit rather than explaining that part of a full
dashboard gap.  Within the retained support, we adapt the classical symmetric
rate--composition decomposition~\cite{kitagawa1955components,dasgupta1978general}
to separate within-video residual differences from video-exposure mix, then
use a stated Shapley convention~\cite{shorrocks2013decomposition}.  We freeze
the same user multiplicities across both logs for bootstrap inference and
across four early/late strata for a temporal audit.  The primary intervals
condition on the realized videos and dates; separate perturbation diagnostics
probe this finite-catalog/calendar boundary.
We organize the evidence as an audit stack: a paired-support discrepancy, a
complete-support error ledger, a frozen prospective calibration, and the
resulting top-$k$ decision loss.  A known-truth semi-synthetic benchmark then
separates the conditions under which user demeaning helps from those it cannot
repair.

Our contributions are fourfold:
\begin{itemize}
  \item a conditional audit protocol that reports the selection funnel and
  separates level, centered shape, rank, and decision discrepancies on paired
  user and video support;
  \item a complete-support, level--shape error ledger that combines support
  entry/exit with Kitagawa-style within-video and video-mix accounting;
  \item a leakage-resistant prospective protocol that fits on history and an
  early exploration window, freezes its parameters, and evaluates a later
  random-insertion reference; and
  \item decision-aware inference and a known-truth stress test exposing where
  level calibration, demeaning, and propensity weighting succeed or fail.
\end{itemize}
The contribution is an audit design rather than a new causal estimator.

\section{Related Work}

\noindent\textbf{Logged-policy evaluation and measurement audits.}
Randomized recommendation logs support unbiased replay evaluation when the
logging design and acceptance rule are known~\cite{li2011unbiasedoffline};
counterfactual learning-to-rank instead uses examination propensities to learn
rankers from biased clicks~\cite{joachims2017unbiasedltr}.  Offline A/B studies
connect counterfactual metrics to online outcomes and candidate decisions
~\cite{gilotte2018offlineab,gruson2019offline}, while large action spaces and
deficient support expose identification--variance tradeoffs
~\cite{saito2022largeactionope,sachdeva2020deficientsupport}.  Measurement
audits cover policy-shift debiasedness, temporal leakage, and agreement with
dense truth~\cite{deffayet2023debiasedness,ji2023dataleakage,
parajuli2026convergent}.  We instead audit a category dashboard's level,
shape, support path, and top-$k$ decisions.

\noindent\textbf{Policy-shaped recommendation logs.}
Serving policies reshape offline populations~\cite{chaney2018algorithmic,
castells2022offline}.  KuaiRec, KuaiRand, and Open Bandit provide dense,
randomized, or multiple-policy logs~\cite{gao2022kuairec,gao2022kuairand,
saito2021openbandit}; short-video position models motivate our unobserved-rank
boundary~\cite{jeunen2023position}.

\noindent\textbf{MNAR learning and evaluation.}
Prior work corrects selected ratings and implicit feedback with inverse
propensities~\cite{schnabel2016recommendations,yang2018unbiased}, proposes
unbiased learning objectives~\cite{saito2020unbiased}, and combines exposure
and outcome models~\cite{wang2019doublyrobust}.  Such procedures require
overlap and identification assumptions~\cite{gao2024causalsurvey}.  We instead
use random insertions as an external measurement reference and make failed
overlap visible; we do not propose a new unbiased learner or policy-value
estimator.

\noindent\textbf{Standardization and decomposition.}
Offline conclusions depend on filtering, targets, aggregation, and scoring
~\cite{canamares2020offline}; common support improves comparability while
changing the target population~\cite{crump2009overlap}.  We adapt classical
rate--composition and Shapley decompositions
~\cite{kitagawa1955components,dasgupta1978general,
shorrocks2013decomposition} to category means and report excluded support
separately.

\noindent\textbf{Category-level measurement.}
RADio shows how references and metrics shape category distributions in
diversity audits~\cite{vrijenhoek2024radio}; our endpoint is instead category
outcome means, excluded support, and downstream top-$k$ decisions.

Unlike unbiased-learning or policy-value work, we audit an analyst-facing
dashboard whose level, centered shape, support, item mix, and top-$k$ loss can
move differently.  The novelty is this estimand-and-diagnostic stack and its
failure-boundary evaluation---not new identification machinery.

\section{Audit Design}
\label{sec:method}

\noindent\textbf{Audit map.}
The method has three passes: compare production and random category vectors on
paired support; account for within-video, exposure-mix, and excluded-support
contributions; then freeze an early calibration before evaluating a later
dashboard and its top-$k$ decisions.  Each pass answers a distinct descriptive
question; none identifies an intrinsic category effect.

\subsection{Conditional paired-support estimand}

Let $Y_e^b\in[0,1]$ be the clipped watch ratio of impression $e$ in log
$b\in\{R,S\}$, and let $c(i)=k$ be video $i$'s official first-level category.
We first restrict both logs to the random-insertion dates, interface tab 1,
valid labels, and initially common videos.  The primary audit requires a
standard user to realize at least five categories over the full window, giving
user demeaning repeated cross-category information.  Five is a coverage design
choice, not an identification condition or a value selected to optimize an
error metric.  Eligible user IDs are intersected with the random-insertion log
and applied to \emph{both} logs; common-video and standard-coverage restrictions
are then iterated to a fixed point $(\mathcal U^*,\mathcal V^*,\mathcal K)$,
with $K=|\mathcal K|$.

The primary contemporaneous estimands below are conditional on this selected
overlap population; the temporal, prospective, and synthetic analyses define
their own cohorts or targets explicitly.  The primary population does not
represent all platform users, and common-video restriction excludes
policy-induced video entry and exit.  Such overlap restriction improves
comparability while changing the target population~\cite{crump2009overlap}.

This is paired clustering, not event-level pairing: a user's random-insertion and
standard rows need not concern the same video or moment.  In bootstrap sample
$t$, one user multiplicity $m_u^{(t)}$ weights all of that user's rows in both
logs.  This preserves cross-log dependence without claiming paired
trials.
The resulting intervals target user-sampling variation conditional on the
selected users' realized videos and dates, not an infinite catalog, future
calendar, or identification uncertainty.

\subsection{Level, shape, rank, and aggregation targets}

For user $u$ and category $k$, let $n^b_{uk}$ and $T^b_{uk}$ be the number and
outcome sum of impressions in log $b$.  The random-insertion reference and
naive standard mean are both impression-row weighted:
\begin{align}
r_k&=\frac{\sum_uT^R_{uk}}{\sum_un^R_{uk}}, &
s_k^{\mathrm{row}}&=\frac{\sum_uT^S_{uk}}{\sum_un^S_{uk}}. \label{eq:rowmeans}
\end{align}
We write $s_k=s_k^{\mathrm{row}}$ when no aggregation superscript is shown.
Let $w_k=\sum_un^R_{uk}/\sum_j\sum_un^R_{uj}$ be fixed random-insertion
category shares.  For any vector $a$, define
$\mathcal C_w(a)_k=a_k-\sum_jw_ja_j$.  We report
\begin{align}
 \mathrm{MAE}_{\mathrm{abs}}(a,r) &= K^{-1}\sum_k |a_k-r_k|,\\
 \mathrm{MAE}_{\mathrm{shape}} &= K^{-1}\sum_k
 |\mathcal C_w(a)_k-\mathcal C_w(r)_k|. \label{eq:metrics}
\end{align}
Categories receive equal weight in these headline MAEs.  We also report
random-share-weighted error, Spearman correlation, maximum error
$\max_k|a_k-r_k|$, and top-10 set overlap
$|T_{10}(a)\cap T_{10}(r)|/10$.

Our cell-equal (CE) and row-weighted (RW) demeaning diagnostics share a user
residual but target different aggregations.  Let
$\mu_u^S=\sum_jT^S_{uj}/\sum_jn^S_{uj}$.  CE gives every
observed user--category cell equal weight; its residual and
random-level-anchored vector are
\begin{align}
d_k^{\mathrm{CE}}&=\frac{\sum_{u:n^S_{uk}>0}
(T^S_{uk}/n^S_{uk}-\mu_u^S)}{|\{u:n^S_{uk}>0\}|}, &
\widehat s_k^{\mathrm{CE}}&=d_k^{\mathrm{CE}}+\bar r, \label{eq:ce}
\end{align}
where $\bar r=\sum_kw_kr_k$.  RW instead preserves impression-row weights,
$d_k^{\mathrm{RW}}=\sum_u(T^S_{uk}-n^S_{uk}\mu_u^S)/\sum_un^S_{uk}$.
Thus CE changes user weighting as well as removing user level.  Its residual
vector is computable from standard impressions alone, but its reported error
is evaluated against the random-insertion reference; absolute error also
borrows $\bar r$.  It is therefore a diagnostic, not a deployable correction.

\subsection{What user demeaning can remove}

The informal statement that demeaning ``removes user effects'' needs a sharper
qualification.  Let the noise-free standard cell mean be
$m^S_{uk}=\alpha_u+\tau_k+\gamma_{uk}$ and let
$\pi^S_{uj}=n^S_{uj}/\sum_\ell n^S_{u\ell}$ denote user $u$'s realized category
portfolio.  Write $E_k[\cdot]$ for an equal average over users with
$n^S_{uk}>0$.  Direct substitution into Eq.~\eqref{eq:ce} gives
\begin{align}
d_k^{\mathrm{CE}}
 &=\tau_k-B_k+G_k, \label{eq:cebias}\\
B_k&=E_k\!\left[\sum_j\pi^S_{uj}\tau_j\right], &
G_k&=E_k\!\left[\gamma_{uk}-\sum_j\pi^S_{uj}\gamma_{uj}\right].
\end{align}
Thus $\alpha_u$ cancels exactly, but realized portfolio $B_k$ and
user--category interaction $G_k$ remain.  Even when $\gamma_{uk}=0$, CE
recovers centered category shape only if $B_k$ is constant across $k$.  RW has the same form
with users weighted by $n^S_{uk}$.  This identity explains why demeaning is a
useful diagnostic under level selection but not a generally unbiased category
estimator; the condition is examined rather than presumed in
Section~\ref{sec:benchmark}.

\subsection{Kitagawa-style accounting for video mix}

Exact video support does not imply equal video weights.  Within category $k$,
let $q^b_{ik}$ be video $i$'s exposure share and $\bar Y^b_{ik}$ its mean in
log $b$.  The observed gap $D_k=s_k-r_k$ obeys
\begin{align}
W_k &= \tfrac12\sum_i(q^S_{ik}+q^R_{ik})
                 (\bar Y^S_{ik}-\bar Y^R_{ik}),\\
M_k &= \tfrac12\sum_i(q^S_{ik}-q^R_{ik})
                 (\bar Y^S_{ik}+\bar Y^R_{ik}),\\
D_k &= W_k+M_k. \label{eq:decomp}
\end{align}
$W_k$ is a symmetric within-video residual difference and $M_k$ a symmetric
video-exposure-mix component.  This is a direct adaptation of classical
rate--composition decomposition~\cite{kitagawa1955components}.  We allocate
$|D_k|$ under a two-player Shapley convention~\cite{shorrocks2013decomposition}, e.g.,
$\phi^W_k=\frac12(|W_k|+|D_k|-|M_k|)$; the two allocations sum exactly to
$|D_k|$ and are nonnegative by the triangle inequality.  Opposing signs can
assign zero to the smaller component.  We repeat the allocation after applying
$\mathcal C_w$ separately to $W$ and $M$.  None of these quantities is a
causal mechanism effect, and the shares depend on the chosen value function.

\subsection{From common support to a complete error ledger}

Let superscripts $F$ and $C$ denote bucket-specific full video support and
common video support for the same paired users.  The full dashboard gap admits
the exact path
\begin{align}
D_k^F=s_k^F-r_k^F
  &=\underbrace{(s_k^F-s_k^C)}_{S_k^S}
    +\underbrace{W_k+M_k}_{s_k^C-r_k^C}
    +\underbrace{(r_k^C-r_k^F)}_{S_k^R}. \label{eq:fullpath}
\end{align}
The two support terms include both exclusive-video rows and the induced
renormalization of retained-video weights.  This restores the part intentionally
excluded by the paired common-support estimand.

To keep level and shape additive, use full-random category shares $w^F$, define
$\bar x^F=\sum_k w_k^F x_k$, $\mathcal C_{w^F}(x)=x-\bar x^F\mathbf 1$, and
$\langle x,z\rangle_{w^F}=\sum_k w_k^F x_kz_k$.  For components
$g_j\in\{S^S,W,M,S^R\}$, define
\begin{align}
A_j^{\mathrm{level}}&=\bar g_j^F\bar D^F, &
A_j^{\mathrm{shape}}&=\langle\mathcal C_{w^F}(g_j),
                           \mathcal C_{w^F}(D^F)\rangle_{w^F}. \label{eq:ledger}
\end{align}
Then $\sum_jA_j^{\mathrm{level}}=(\bar D^F)^2$ and
$\sum_jA_j^{\mathrm{shape}}=\|\mathcal C_{w^F}(D^F)\|_{w^F}^2$ exactly.  Contributions
are signed: cancellation is information, not an error to clip away.  This is
an application of inner-product bookkeeping rather than a novel identity or
causal attribution.

\subsection{Prospective calibration and decision loss}

The contemporaneous CE diagnostic in Eq.~\eqref{eq:ce} uses the evaluation
window.  For a prospective audit we instead estimate user baselines from
historical standard traffic, estimate either a scalar gap $\delta$ or category
gap vector $\boldsymbol\delta$ from an early random/standard exploration
window, freeze both support and parameters, and apply
$\widehat{\boldsymbol s}_{\rm test}+\delta\mathbf 1$ or
$\widehat{\boldsymbol s}_{\rm test}+\boldsymbol\delta$ in a later window.
For every deployable estimator, test-random outcomes are used only as the
evaluation reference.  We separately show an explicitly non-deployable oracle
that uses the test-random grand mean.  This evaluates temporal transport of a
calibration map, not an intervention effect.

For a size-$k$ category decision, define plug-in reference regret
\begin{align}
\mathcal R_k(a;r)=k^{-1}\!\sum_{j\in T_k(r)}r_j
                 -k^{-1}\!\sum_{j\in T_k(a)}r_j, \label{eq:regret}
\end{align}
which is nonnegative and complements set overlap.  A simple sufficient ranking
certificate is
$\|\mathcal C_w(a-r)\|_\infty<\frac12(r_{(k)}-r_{(k+1)})$; failure of this
conservative condition does not imply an incorrect set.  We report paired
bootstrap regret differences and category inclusion probabilities.  We also
report simultaneous bootstrap max-$t$ signed-error bands over the 38
categories~\cite{dudoit2004singlestep}; these bands do not adjust the separate
family of cutoff-by-estimator regret contrasts.

\section{Data and Evaluation}
\label{sec:data}

\noindent\textbf{KuaiRand.}
Randomization is within recommendation-list events, not across disjoint user
traffic arms.  At a list request, an intervention is triggered with a fixed
but undisclosed probability; conditional on intervention, one listed video is
chosen at random and replaced by an item sampled uniformly from a 7,583-video
pool formed after filtering spam such as advertisements
~\cite{gao2022kuairand}.  KuaiRand-Pure retains only interactions involving
that pool, so its standard-serving file is a candidate-pool slice, not the full
production feed.  List position and the replaced item are unavailable.  This
prevents a position-aware feed analysis of the kind enabled by explicit rank
interventions~\cite{jeunen2023position}.

The base audit uses April 22--May 8, 2022, tab 1, and the within-KuaiRand
intersection of \BaseVideos{} videos in \BaseCategories{} categories.  It
contains \BaseRandomRows{} random-insertion rows from \BaseRandomUsers{} users and
\BaseStandardRows{} standard rows from \BaseStandardUsers{} users.  The
five-category rule requires repeated cross-category exposure while retaining
roughly half of base standard users; it keeps \EligibleStandardUsers{} users
(\EligibleStandardUserShare\% of the base standard users); five are absent
from random insertions.  The fixed point then contains \PairedUsers{} identical
user IDs (\PairedBaseUserShare\% of base standard users), \PairedVideos{}
identical video IDs (\PairedBaseVideoShare\% of the initial common videos),
\PairedRandomRows{} random-insertion rows, and \PairedStandardRows{} standard
rows.  This funnel is part of the estimand, not a claim of population-wide
coverage.

Categories come from KuaiRand's official supplementary mapping, joined by
\texttt{final\_video\_id}; the two datasets' independently re-indexed IDs are
never crossed.  The supplementary first-level labels carry prediction
probabilities and are not manually adjudicated truth.  Twelve of 7,583 Pure
video IDs lack a valid first-level label.  A sensitivity analysis rebuilds the
scope after requiring probability at least 0.8.

The primary paired bootstrap uses \MainBootstrapDraws{} synchronized
user-cluster draws while holding videos and dates fixed.  For temporal
stability we predefine April 22--30 and
May 1--8, then simultaneously iterate
user and video intersections across random/standard $\times$ early/late.
Full-period eligibility is frozen; we do not reselect users by realized
half-period category coverage.  One bootstrap multiplicity weights all four
strata in \TemporalBootstrapDraws{} temporal draws.  Click and long-view
indicators are evaluated on the same primary scope.  Here click is the
released \texttt{is\_click} indicator; long view is the released
\texttt{long\_view} label from KuaiRand's watch-time thresholding.

\noindent\textbf{Prospective split.}
For the out-of-time experiment, April 8--21 standard traffic determines the
five-category eligibility rule and historical user baselines.  April 22--30
random/standard traffic estimates frozen scalar and category calibration maps;
May 1--8 is untouched until evaluation.  Fit-time common users and videos are
iterated across history standard and both calibration logs.  The final test
estimand is explicitly conditional on frozen users who return in both test
logs on frozen video support.  Exposure presence, but not test outcomes,
therefore enters evaluation-cohort selection; the result cannot be transported
to absent users without assumptions.

\noindent\textbf{Known-truth benchmark.}
To probe assumptions unavailable in the logs, we simulate
\BenchmarkReplications{} finite worlds with 38 categories, 760 videos,
1,000 equal-weight users, and \BenchmarkRowsPerReplication{} Bernoulli
impressions per world.  Target category shares equal the paired-random shares;
the target averages uniformly over users and videos within category.  Serving
selects user level, user--category affinity, video quality, and overlap in
controlled scenarios.  Because the complete conditional-mean surface and
serving probabilities are known, we compare naive, scalar-anchor, CE/RW, and
oracle H\'ajek diagnostics against exact truth.  Across-world percentile ranges
describe stress-test heterogeneity, not confidence intervals for KuaiRand.

\section{Paired KuaiRand Audit}
\label{sec:kuairand}

\subsection{A large level gap but moderate shape gap}

\begin{figure*}[t]
  \centering
  \includegraphics[width=0.94\textwidth]{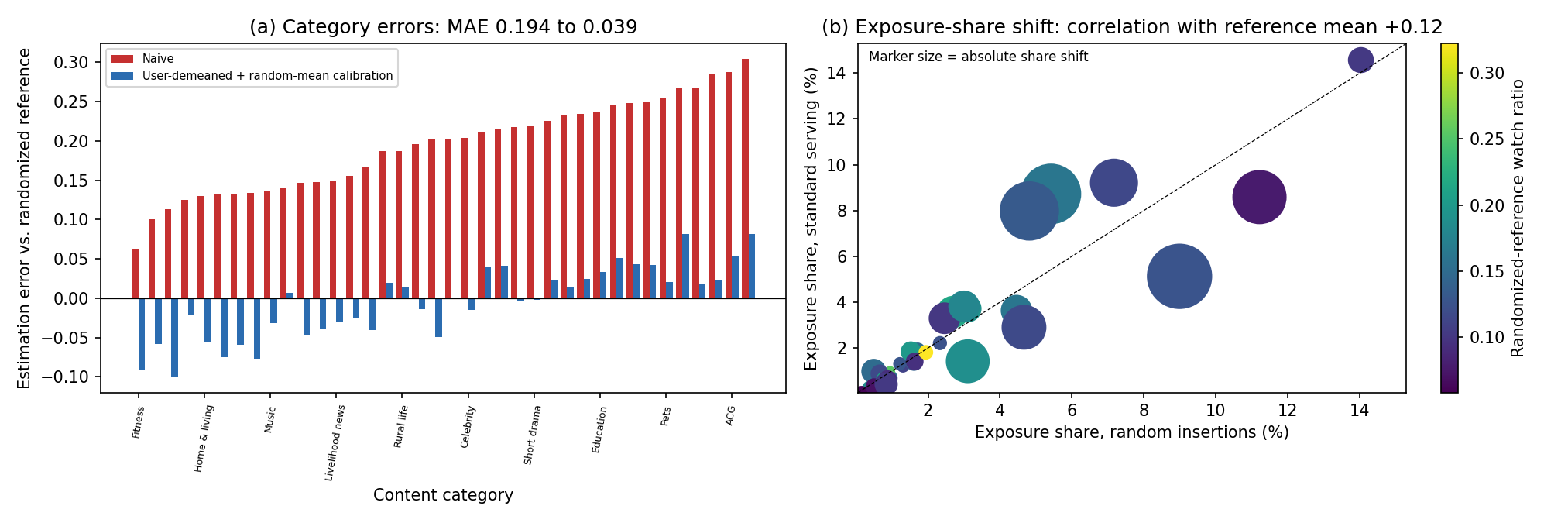}
  \Description{Two panels show paired-scope category errors relative to the
  random-insertion reference and random versus standard category exposure shares.}
  \caption{Paired-user, exact-video-support audit.  Left: standard category
  errors and random-level-anchored CE errors.  Right: random-insertion
  versus standard category exposure shares.  Panel (b)'s title reports the Pearson
  correlation between the standard-minus-random share shift and the reference
  category mean.  The random-insertion log is a reference for the realized
  category--video mix, not an intrinsic category effect.}
  \label{fig:bias}
\end{figure*}

The paired standard grand mean is \StandardGrandMean{}, versus
\RandomGrandMean{} in random-insertion impressions.  Accordingly, naive category means
have absolute MAE \NaiveAbs{} and maximum error \NaiveMaximum{}.  Centering
both vectors with random exposure shares reduces MAE to \NaiveCentered{}.
Rank correlation is already \NaiveRank{}, and \NaiveTopTen{} of the reference
top 10 are retained.  The discrepancy is therefore primarily level, not a
wholesale rank reversal.  Figure~\ref{fig:bias} visualizes the category errors
and exposure-share shifts.

Observed-cell-equal (CE) demeaning followed by the random grand-mean anchor
yields absolute MAE \DemeanAbs{}, an \AbsoluteReduction\% reduction.  Its
reference-evaluated centered MAE improves by only \CenteredReduction\%, from
\NaiveCentered{} to \DemeanCentered{}; rank changes to \DemeanRank{} and
top-10 overlap to \DemeanTopTen{}.  The paired bootstrap gives direct CE-minus-
naive changes of \AbsDifference{} \AbsDifferenceCI{} for absolute MAE,
\CenteredDifference{} \CenteredDifferenceCI{} for centered MAE, and
\RankDifference{} \RankDifferenceCI{} for rank.  Hence centered improvement is
detectable on this scope, but rank improvement is not.  These intervals
quantify user sampling variation conditional on fixed videos, observed dates,
and the selected scope, not identification uncertainty.

Restricting the reference from all random users to the paired users changes
category means by only 0.0041 MAE (rank correlation 0.9954, top-10 overlap
1.0) on the same final videos.  Thus cohort pairing removes an important design
ambiguity without driving the substantive result.

The more consequential restriction is video support.  On the same users but
with each log's full valid video support, naive absolute/centered MAE and rank
are \FullSupportAbs{}/\FullSupportCentered{}/\FullSupportRank{}; exact common
support yields \NaiveAbs{}/\NaiveCentered{}/\NaiveRank{}.  Common support keeps
\CommonRandomRowShare\% of random-insertion rows but \CommonStandardRowShare\%
of standard rows (\FullRandomVideos{}/\FullStandardVideos{} full-support videos
versus \CommonSupportVideos{} common videos), so it excludes support substitution
and modestly improves alignment.  Table~\ref{tab:scope} exposes the threshold
and support choices rather than treating five categories as canonical.
Rebuilding the fixed point from threshold 1 to 10 reduces the cohort from
\ThresholdUsersOne{} to \ThresholdUsersTen{} users, while naive and CE absolute
MAE remain within \ThresholdNaiveAbsRange{} and \ThresholdDemeanAbsRange{},
respectively.  Thus the headline level pattern is stable, although the target
population changes sharply.
Row-weighted demeaning gives absolute/centered MAE
\RowDemeanAbs{}/\RowDemeanCentered{} and rank \RowDemeanRank{}, close to CE;
the finding is not an artifact of cell-equal aggregation alone.  If categories
are instead weighted by random-insertion shares, naive absolute/centered MAE is
\WeightedNaiveAbs{}/\WeightedNaiveCentered{} and CE MAE is
\WeightedDemeanAbs{}/\WeightedDemeanCentered{}.

\begin{table*}[t]
  \centering
  \small
  \caption{Selection sensitivity.  Every threshold rebuilds the paired fixed
  point; metrics are descriptive point estimates.}
  \label{tab:scope}
  \begin{tabular}{lrrrrr}
    \toprule
    Min. categories & 1 & 2 & 3 & 5 & 10 \\
    \midrule
    Paired users & \ThresholdUsersOne & \ThresholdUsersTwo & \ThresholdUsersThree & \ThresholdUsersFive & \ThresholdUsersTen \\
    Naive abs. MAE & \ThresholdNaiveAbsOne & \ThresholdNaiveAbsTwo & \ThresholdNaiveAbsThree & \ThresholdNaiveAbsFive & \ThresholdNaiveAbsTen \\
    Naive centered MAE & \ThresholdNaiveCenteredOne & \ThresholdNaiveCenteredTwo & \ThresholdNaiveCenteredThree & \ThresholdNaiveCenteredFive & \ThresholdNaiveCenteredTen \\
    Naive rank $\rho$ & \ThresholdNaiveRankOne & \ThresholdNaiveRankTwo & \ThresholdNaiveRankThree & \ThresholdNaiveRankFive & \ThresholdNaiveRankTen \\
    CE abs. MAE & \ThresholdDemeanAbsOne & \ThresholdDemeanAbsTwo & \ThresholdDemeanAbsThree & \ThresholdDemeanAbsFive & \ThresholdDemeanAbsTen \\
    \bottomrule
  \end{tabular}
\end{table*}

\subsection{Absolute and centered video-mix accounting}

Equation~\eqref{eq:decomp} reconstructs every category gap to numerical error
below \AccountingIdentityError{}.  Across categories, mean $|W_k|$ is
\WithinVideoMeanAbs{} and mean $|M_k|$ is \ItemMixMeanAbs{};
$|W_k|>|M_k|$ in \WithinVideoDominatesCount{} of
\AccountingCategoryCount{} categories.  Because component
MAEs are not additive under cancellation, our headline shares use exact
Shapley accounting: the mean absolute gap \NaiveAbs{} is allocated as
\ResponseShapley{} (\ResponseShapleyShare\%) to the within-video residual and
\ItemMixShapley{} (\ItemMixShapleyShare\%) to video mix.  After fixed-share
centering, the \NaiveCentered{} gap is instead allocated as
\CenteredResidualShapley{} (\CenteredResidualShare\%) and
\CenteredItemMixShapley{} (\CenteredItemMixShare\%), respectively.  Thus video
mix is a minority of the absolute gap but roughly half of category shape.
The \MainShapleyBootstrapDraws{}-draw paired-user intervals use one shared
user-level positive exponential multiplier in both logs.  Intervals for the absolute
shares are \AbsoluteResidualSharePctCI{}\% and \AbsoluteItemMixSharePctCI{}\%;
centered-share intervals are \CenteredResidualSharePctCI{}\% and
\CenteredItemMixSharePctCI{}\%.

This result sharpens, but does not causalize, the audit.  Matching video IDs
removes support substitution from this estimand; video mix shows what unequal
weights within that support still contribute.  The within-video residual can
contain audience weighting, state, position, time, or other unobserved log
differences.

\subsection{Complete-support level--shape ledger}

\begin{figure*}[t]
  \centering
  \includegraphics[width=0.96\textwidth]{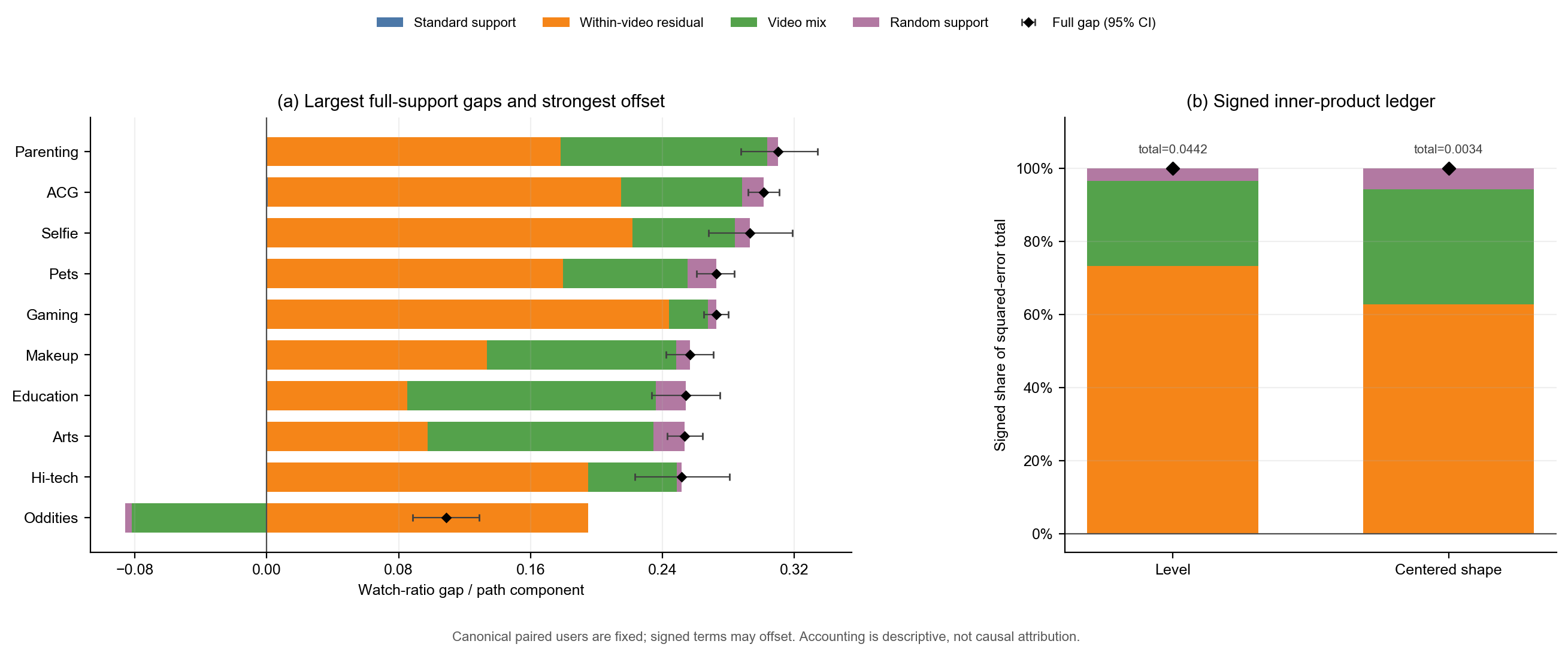}
  \Description{A signed stacked-bar path reconstructs selected full-support
  category gaps. A second panel allocates level and centered-shape squared
  errors to support, within-video residual, and video mix.}
  \caption{Complete dashboard-error ledger.  The same paired users are held
  fixed while bucket-specific video support is restored.  Panel (b) reports
  signed inner-product shares; these are exact accounting, not causal effects.
  Black whiskers resample users with positive multipliers while holding the
  realized videos and dates fixed.}
  \label{fig:ledger}
\end{figure*}

The common-support audit omits \LedgerRandomOnlyVideos{} random-only and
\LedgerStandardOnlyVideos{} standard-only videos.  Restoring bucket-specific
support for the same \PairedUsers{} users and applying Eq.~\eqref{eq:fullpath}
reconstructs every full-support category gap to machine precision.  Under
full-random exposure weights, the mean gap is \LedgerFullWeightedMeanGap{} and
weighted root mean squared error (RMSE) is \LedgerWeightedRMSE{}.
Equation~\eqref{eq:ledger} allocates
the total squared error as \LedgerStandardSupportTotalShare\% standard-support,
\LedgerWithinTotalShare\% within-video residual, \LedgerMixTotalShare\% video
mix, and \LedgerRandomSupportTotalShare\% random-support contributions.  The
\LedgerBootstrapDraws{}-draw paired positive-multiplier intervals are
\LedgerWithinTotalShareCI\% and \LedgerMixTotalShareCI\% for the two dominant
shares.  Strictly positive multipliers avoid dropping a fixed-support video
from any resample.
Figure~\ref{fig:ledger} visualizes this complete-support accounting path.

The split changes for shape: the contributions are
\LedgerStandardSupportShapeShare\%
standard support, \LedgerWithinShapeShare\% within-video,
\LedgerMixShapeShare\% video mix, and \LedgerRandomSupportShapeShare\% random
support.  Thus the support path contributes a small but nonzero fraction of the
full discrepancy, while video mix matters more for shape than for total squared
error.  Signed inner products permit components to offset; shares are
descriptive allocations under a fixed path.

\begin{figure*}[t]
  \centering
  \includegraphics[width=0.96\textwidth]{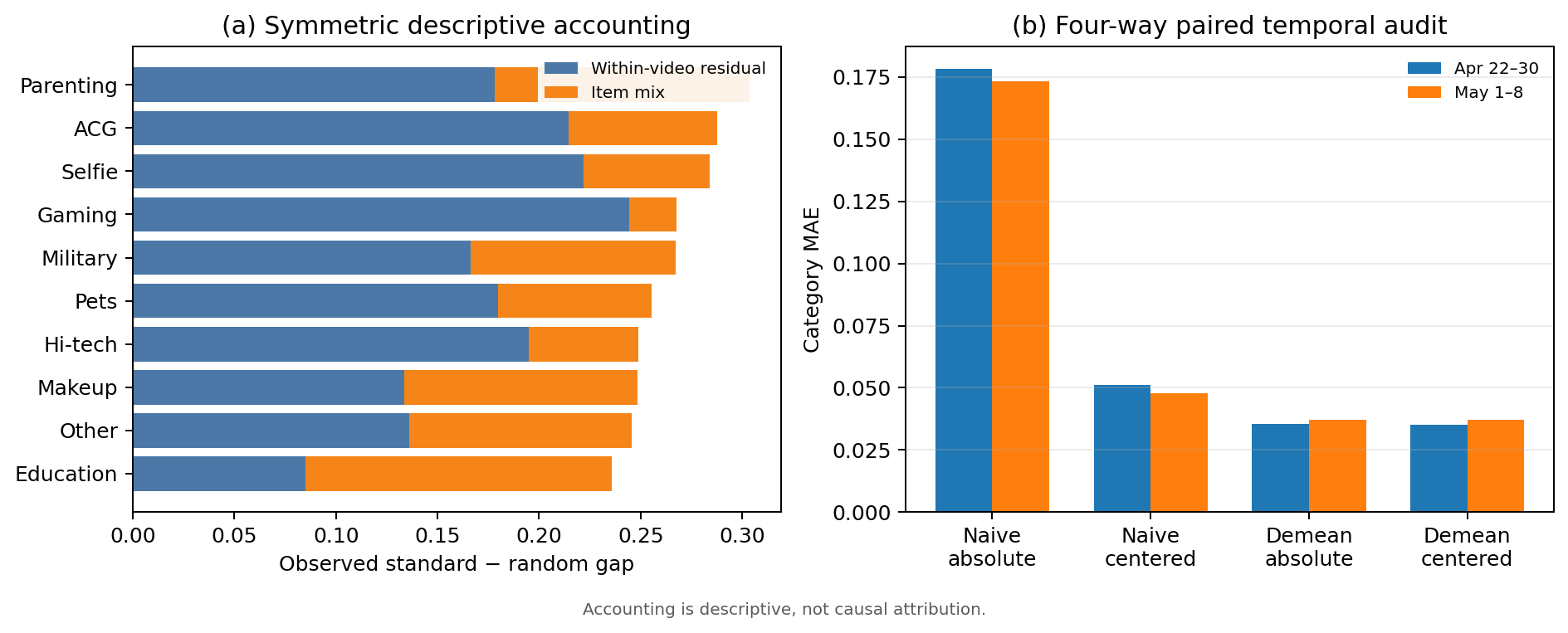}
  \Description{The left panel decomposes the ten largest category gaps into
  symmetric within-video-residual and video-mix terms.  The right panel compares
  early and late category errors on four-way common support.}
  \caption{Descriptive gap accounting and temporal stability.  Panel (a)
  displays the exact signed identity, not mechanism effects.  Panel (b) fixes
  \TemporalUsers{} users and \TemporalVideos{} videos in all four log--time
  strata; no half-period eligibility filter is applied.}
  \label{fig:paired}
\end{figure*}

\subsection{Temporal, outcome, and label stress tests}

The four-way support in Figure~\ref{fig:paired} contains \TemporalUsers{} users
and \TemporalVideos{} videos in every stratum.  Early versus late naive absolute MAE is
\EarlyNaiveAbs{} versus \LateNaiveAbs{}, centered MAE is
\EarlyNaiveCentered{} versus \LateNaiveCentered{}, and rank correlation is
\EarlyNaiveRank{} versus \LateNaiveRank{}.  Joint bootstrap differences are
\TemporalAbsChange{} \TemporalAbsChangeCI{}, \TemporalCenteredChange{}
\TemporalCenteredChangeCI{}, and \TemporalRankChange{}
\TemporalRankChangeCI{}, respectively.  All intervals include zero.  The
naive category-wise gap vectors remain correlated at \TemporalGapRank{}
\TemporalGapRankCI{}.  We find no clear temporal drift over this short window.

A finite-calendar/catalog sensitivity holds the selected scope fixed.  Jointly
deleting each of \DependenceDateOmissions{} dates preserves lower absolute,
centered, and maximum error in every deletion; the centered CE-minus-naive
change ranges over \DependenceDateCenteredChangeRange{} (a
\DependenceDateCenteredReductionRange{} reduction).  Under
\DependenceVideoDraws{} shared positive video-weight perturbations, centered
error improves in \DependenceVideoCenteredFavorableShare\% of draws and its
2.5--97.5\% sensitivity envelope is
\DependenceVideoCenteredEnvelope{}.  These are deletion/perturbation ranges,
not calibrated multiway confidence intervals.  Rank and top-10 improve in only
\DependenceVideoRankFavorableShare\% and
\DependenceVideoTopTenFavorableShare\% of video perturbations, reinforcing the
absence of a stable ordering claim.

\begin{table*}[t]
  \centering
  \small
  \caption{Same-scope outcome sensitivity.  ``CE demean + R mean'' borrows each
  outcome's random-insertion grand-mean level.  Ranking alignment is outcome
  specific and does not consistently improve after demeaning.}
  \label{tab:outcomes}
  \begin{tabular}{llrrrrr}
    \toprule
    Outcome & Estimate & Abs. MAE & Centered MAE & Max error & Rank $\rho$ & Top-10 \\
    \midrule
    Watch ratio & Naive & \NaiveAbs & \NaiveCentered & \NaiveMaximum & \NaiveRank & \NaiveTopTen \\
    Watch ratio & CE demean + R mean & \DemeanAbs & \DemeanCentered & \DemeanMaximum & \DemeanRank & \DemeanTopTen \\
    Click & Naive & \ClickNaiveAbs & \ClickNaiveCentered & \ClickNaiveMaximum & \ClickNaiveRank & \ClickNaiveTopTen \\
    Click & CE demean + R mean & \ClickDemeanAbs & \ClickDemeanCentered & \ClickDemeanMaximum & \ClickDemeanRank & \ClickDemeanTopTen \\
    Long view & Naive & \LongNaiveAbs & \LongNaiveCentered & \LongNaiveMaximum & \LongNaiveRank & \LongNaiveTopTen \\
    Long view & CE demean + R mean & \LongDemeanAbs & \LongDemeanCentered & \LongDemeanMaximum & \LongDemeanRank & \LongDemeanTopTen \\
    \bottomrule
  \end{tabular}
\end{table*}

Table~\ref{tab:outcomes} shows that click and long view sharpen the boundary of
the watch-ratio result.
Outcome-specific random anchors greatly reduce absolute MAE, but ordering does
not improve: for click, rank/top-10 changes from
\ClickNaiveRank{}/\ClickNaiveTopTen{} to
\ClickDemeanRank{}/\ClickDemeanTopTen{}; for long view it changes from
\LongNaiveRank{}/\LongNaiveTopTen{} to
\LongDemeanRank{}/\LongDemeanTopTen{}.  The cross-outcome conclusion is that
level calibration can look strong while category ordering remains outcome
specific---not that demeaning recovers a common ranking.  Requiring official
category confidence at least 0.8 retains \ConfidenceUsers{} paired users,
\ConfidenceVideos{} videos, and all 38 categories.  Watch-ratio naive
absolute/centered MAE becomes \ConfidenceNaiveAbs{}/\ConfidenceNaiveCentered{}
with rank \ConfidenceNaiveRank{}; calibrated demeaning gives
\ConfidenceDemeanAbs{}/\ConfidenceDemeanCentered{} with rank
\ConfidenceDemeanRank{}.  The main level/shape pattern is descriptively similar,
although fine ranking remains sensitive.  Without clipping watch ratio to
$[0,1]$, naive absolute/centered MAE and rank are
\UnclippedNaiveAbs{}/\UnclippedNaiveCentered{}/\UnclippedNaiveRank{}; CE gives
\UnclippedDemeanAbs{}/\UnclippedDemeanCentered{}/\UnclippedDemeanRank{}.

\section{Prospective and Decision Audits}
\label{sec:prospective}

\subsection{Does a frozen calibration transport?}

\begin{figure*}[t]
  \centering
  \includegraphics[width=0.91\textwidth]{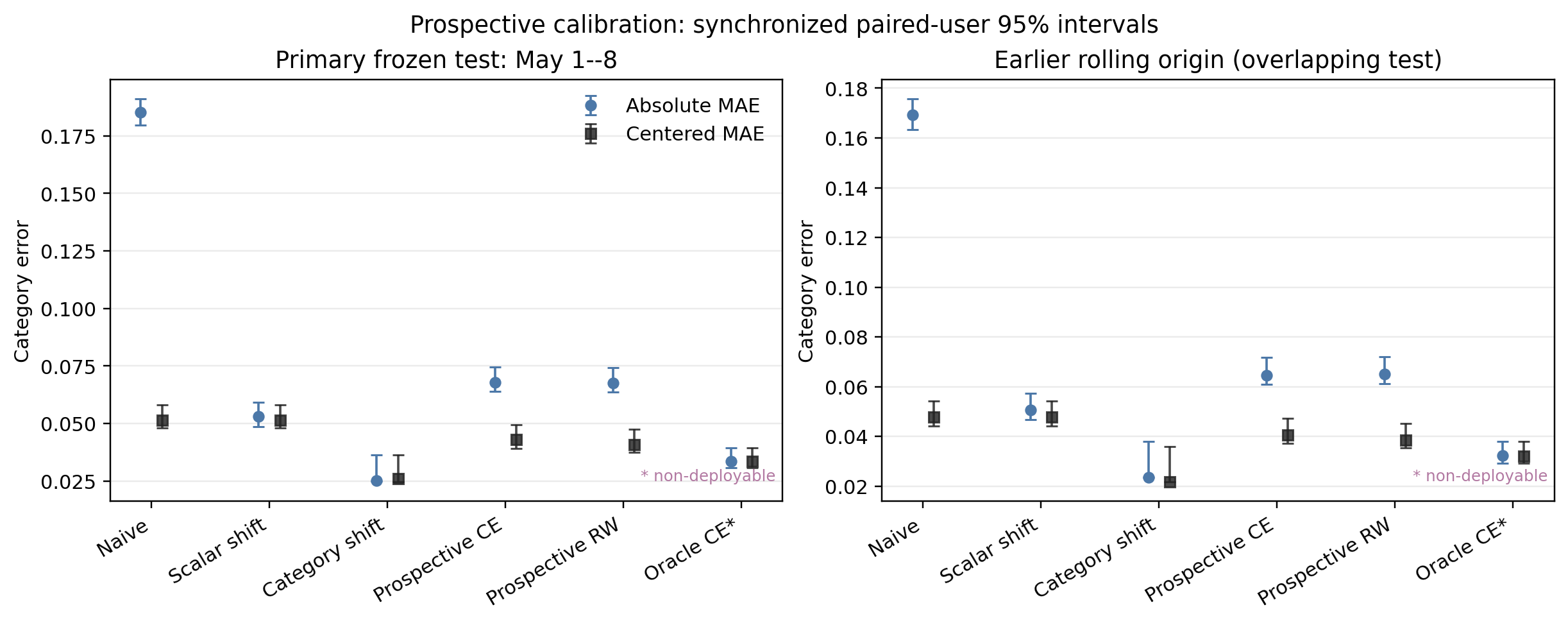}
  \Description{Two panels compare prospective category errors for a primary
  frozen test and an earlier overlapping rolling-origin sensitivity.}
  \caption{Leakage-resistant prospective calibration.  Category shifts are
  learned only from the early exploration window.  Oracle CE uses test-period
  information and is shown only as a non-deployable diagnostic.  The secondary
  rolling origin has an overlapping test and is not an independent replication.
  Intervals resample the frozen fit-time user cohort with one shared user
  multiplicity across all five strata while holding allowed support and window
  dates fixed.}
  \label{fig:prospective}
\end{figure*}

History-only eligibility yields \ProspectiveHistoryEligibleUsers{} users.  The
fit-time fixed point freezes \ProspectiveFrozenUsers{} users and
\ProspectiveFrozenVideos{} allowed videos;
the later returning paired-exposure cohort contains \ProspectiveTestUsers{}
users, or \ProspectiveFrozenRetention\% of frozen and
\ProspectiveHistoryRetention\% of history-eligible users.  Within the frozen
allowed set, realized test support is \ProspectiveTestRandomVideos{} random and
\ProspectiveTestStandardVideos{} standard videos; the prospective protocol allows
test-time support and mixture to differ rather than imposing exact realized-
video matching.  Fit-time allowed support, user baselines, and calibration
shifts are frozen before May 1--8 outcomes are opened; returning users and
realized test support are still determined by test exposure presence.  The
bootstrap samples all \ProspectiveFrozenUsers{} frozen fit-time users, not only
the returning test cohort.  One user multiplicity is synchronized across
history standard, calibration random/standard, and test random/standard in each
of \ProspectiveBootstrapDraws{} draws, recomputing calibration summaries within
draw while retaining the frozen support and per-user history baselines.

Figure~\ref{fig:prospective} contrasts the primary frozen test with the
overlapping rolling-origin sensitivity.  On test, naive absolute MAE is
\ProspectiveNaiveAbs{}.  An early-window scalar
shift lowers it to \ProspectiveScalarAbs{} but leaves centered MAE at
\ProspectiveScalarCentered{} by construction.  Transporting the early
category-specific gap instead yields absolute/centered MAE
\ProspectiveCategoryAbs{}/\ProspectiveCategoryCentered{}, rank
\ProspectiveCategoryRank{}, and top-10 overlap \ProspectiveCategoryTopTen{}.
Its direct differences from scalar calibration are
\ProspectiveCategoryMinusScalarAbs{} \ProspectiveCategoryMinusScalarAbsCI{} for
absolute MAE and \ProspectiveCategoryMinusScalarCentered{}
\ProspectiveCategoryMinusScalarCenteredCI{} for centered MAE.  The early and
late gap vectors correlate at \ProspectiveGapRank{}, which is consistent with
the observed short-horizon transport.

Prospective RW user residualization has absolute/centered MAE
\ProspectiveRWAbs{}/\ProspectiveRWCentered{}: it improves shape over the scalar
map but is worse in absolute level than the category map.  The prospective
result suggests a concrete workflow when recent random exploration is
available: estimate one gap per category, freeze that map, apply it to the next
dashboard, and monitor its out-of-time error rather than correcting only the
grand mean.  Here this map outperforms the scalar anchor over one week; it is a
conditional transport result, not a universal correction, and it does not
cover the 17.2\% of frozen users absent from one test log.

\subsection{Decision value is cutoff-specific}

\begin{figure*}[t]
  \centering
  \includegraphics[width=0.92\textwidth]{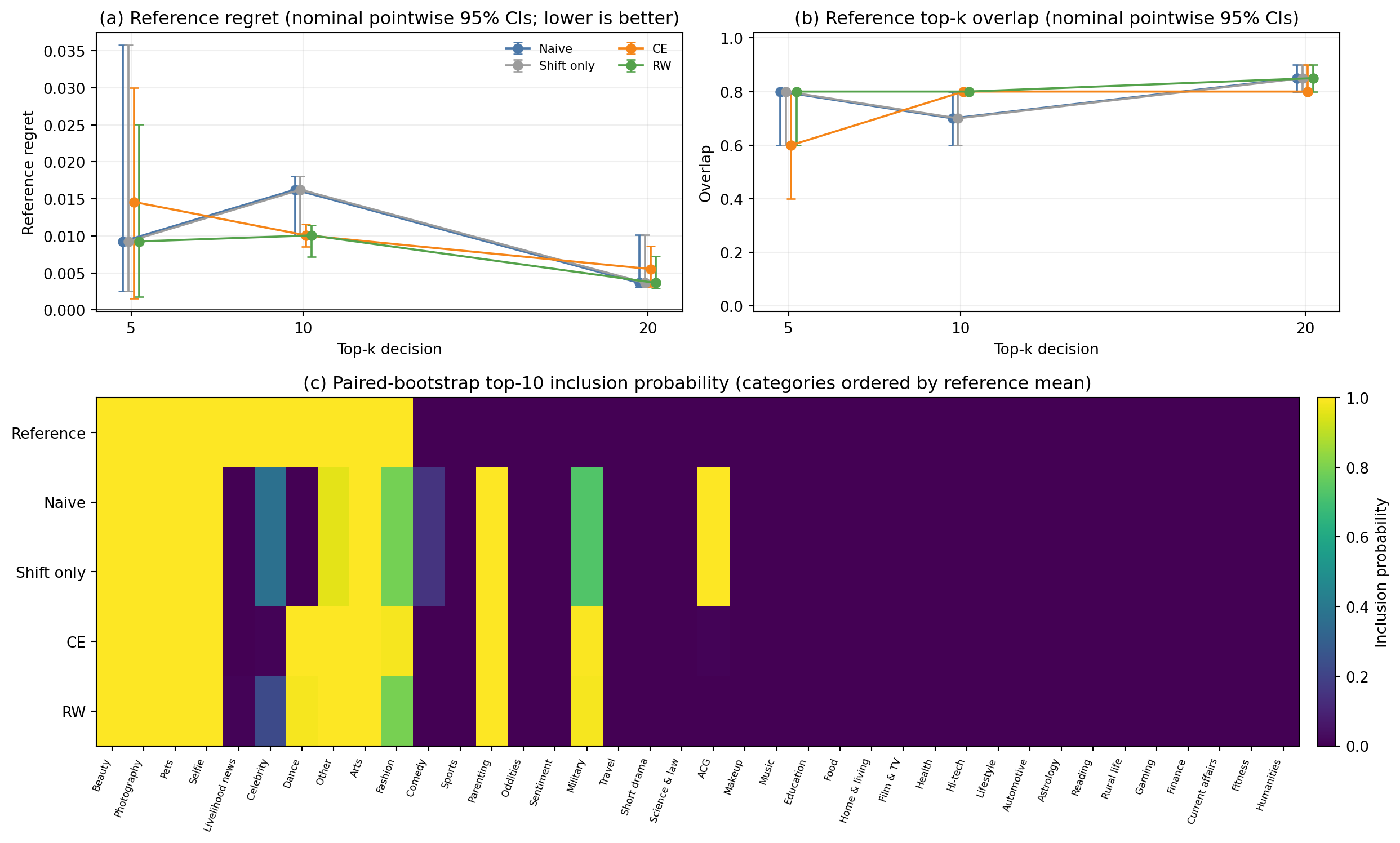}
  \Description{Regret and overlap are shown for top 5, 10, and 20, followed by
  a heat map of paired-bootstrap top-10 category inclusion probabilities.}
  \caption{Decision-aware audit with \DecisionBootstrapDraws{} paired-user
  draws conditional on the realized catalog and calendar.  Error bars are
  nominal percentile intervals.  The exploratory top-10 RW result does not
  generalize to the top-5 or top-20 cutoff.}
  \label{fig:decision}
\end{figure*}

Figure~\ref{fig:decision} reports both regret and overlap.  Reference regret in
Eq.~\eqref{eq:regret} prevents a near-boundary set change
from being treated like a large value loss.  At $k=10$, naive regret is
\DecisionTenNaiveRegret{} \DecisionTenNaiveRegretCI{} with overlap
\DecisionTenNaiveOverlap{}.  RW regret is \DecisionTenRWRegret{}
\DecisionTenRWRegretCI{} with overlap \DecisionTenRWOverlap{}; the paired
RW-minus-naive change is \DecisionTenRWChange{} \DecisionTenRWChangeCI{}.
CE gives the same point regret \DecisionTenCERegret{}, although its direct
paired interval narrowly includes zero.  These are nominal percentile
intervals over six CE/RW-by-cutoff contrasts, not family-wise intervals; the
top-10 exclusion of zero is therefore exploratory.  A single-step standardized
max-$|t|$ interval uses the same \DecisionBootstrapDraws{} draws for the full
six-contrast family~\cite{dudoit2004singlestep} and is
\DecisionTenRWFamilywiseCI{} for RW at $k=10$ (critical value
\DecisionFamilywiseCritical{}) and includes zero; none of the six simultaneous
intervals excludes zero.  A scalar shift is exactly
ranking-invariant in every draw and hence cannot change regret.

To avoid using one noisy random vector both to define and value the reference
set---the selection problem underlying winner's-curse inference
~\cite{andrews2024winners}---we freeze an outcome-free hash split:
\SplitReferenceSelectionRows{}
random rows define top-$k$, while the disjoint
\SplitReferenceEvaluationRows{} rows value every set; standard rows alone rank
the candidate sets.  All \SplitReferencePointDirectionsMaintained{} planned
CE/RW-versus-naive point directions match the original audit, but every nominal
and family-wise interval includes zero.  For RW at $k=10$, the split-reference
contrast is \SplitReferenceTenRWChange{}
\SplitReferenceTenRWChangeCI{} nominally and
\SplitReferenceTenRWFamilywiseCI{} family-wise.  Thus disjoint held-out
reference valuation reinforces the conclusion that no decision gain is
detected.

The benefit is not universal.  At $k=5$, naive regret \DecisionFiveNaiveRegret{}
is smaller than CE's \DecisionFiveCERegret{}, and the RW-minus-naive interval
\DecisionFiveRWChangeCI{} includes zero.  At $k=20$ the corresponding interval
\DecisionTwentyRWChangeCI{} also includes zero.  None of the conservative
margin certificates holds because maximum centered errors exceed half the
reference boundary gaps.  Bootstrap inclusion probabilities expose which
categories cross the top-10 boundary; stability alone is not accuracy, since
an estimator can stably include a nonreference category.  Separate 38-category
simultaneous centered-error bands exclude zero for
\DecisionNaiveCenteredBandExclusions{}/38 naive,
\DecisionCECenteredBandExclusions{}/38 CE, and
\DecisionRWCenteredBandExclusions{}/38 RW errors; they quantify category-wise
misalignment but do not correct the six regret comparisons.

\section{Known-Truth Failure Boundaries}
\label{sec:benchmark}

\begin{figure*}[t]
  \centering
  \includegraphics[width=0.94\textwidth]{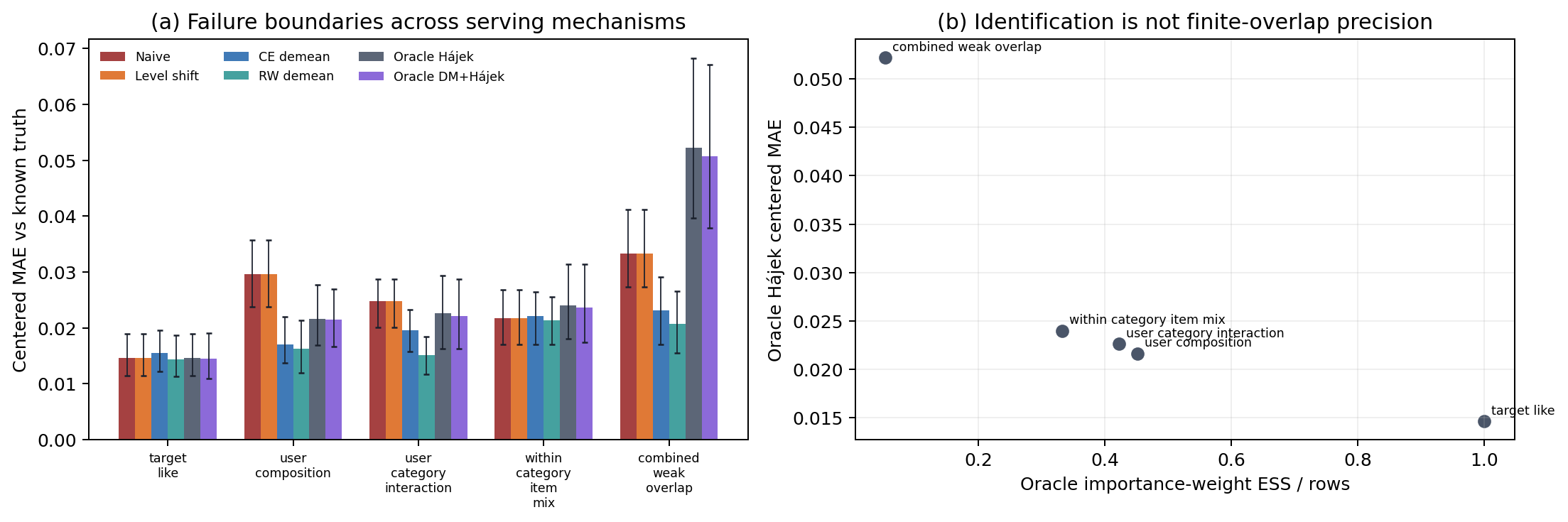}
  \Description{Centered error is compared against exact truth across five
  serving mechanisms; a second panel relates oracle error to importance-weight
  effective sample size.}
  \caption{Known-truth stress test.  Bars are means over
  \BenchmarkReplications{} independently redrawn finite worlds; whiskers are
  across-world 95\% prediction ranges, not KuaiRand confidence intervals.
  Oracle direct-method-augmented H\'ajek (DM+H\'ajek) uses the exact outcome
  surface as a control variate.}
  \label{fig:benchmark}
\end{figure*}

For self-containment, let $u$, $k$, and $i$ index synthetic users, categories,
and videos.  Independent standard-normal scores define clipped centered user
level $a_u$, target-share-centered $\tau_k=0.055z_k$, column-centered
$\gamma_{uk}=\lambda_\gamma z_{uk}$, and within-category-centered
$v_{ik}=0.045z_{ik}$.  The Bernoulli surface is
\begin{equation}
m_{uki}=\operatorname{logit}^{-1}
 \{-0.20+0.24a_u+4\tau_k+4\gamma_{uk}+4v_{ik}\}.
\label{eq:dgp}
\end{equation}
The target shares $q_k$ match paired-random KuaiRand, and target category means
average $m_{uki}$ uniformly over users and videos.  Serving factorizes as
$p_S(k,i\mid u)=p_S(k\mid u)p_S(i\mid k)$: category probabilities exponentially
tilt $q_k$ by user level and affinity, video probabilities tilt the within-class
uniform law by item quality, and each is mixed with its target at a declared
overlap floor.  Scenarios selectively activate these terms.  Known weights give
oracle H\'ajek; DM+H\'ajek uses the exact surface as a control variate, not a
fitted doubly robust learner.  Each replication redraws the world and log, so
Figure~\ref{fig:benchmark}'s whiskers are across-world prediction ranges.

No single adjustment dominates.  Under target-like serving, naive centered MAE
is already \BenchmarkTargetNaiveCentered{}; under logit-additive user selection,
naive/RW error is \BenchmarkUserNaiveCentered{}/\BenchmarkUserRWCentered{}.
With user--category interaction the pair becomes
\BenchmarkInteractionNaiveCentered{}/\BenchmarkInteractionRWCentered{}.
Under pure within-category item selection, RW barely changes error
(\BenchmarkItemNaiveCentered{} to \BenchmarkItemRWCentered{}), as expected:
user demeaning does not rebalance videos.  In a mean-scale exactly additive
positive control, RW reduces centered MAE from
\BenchmarkPositiveNaiveCentered{} to \BenchmarkPositiveRWCentered{}.  The
incomplete reduction is consistent with the remaining portfolio term in
Eq.~\eqref{eq:cebias}; it does not establish unbiased shape recovery.

Known propensities do not guarantee precision.  Under combined weak overlap,
global importance-weight effective sample size (ESS) is 5.3\% of rows, mean
minimum category ESS is \BenchmarkWeakMinCategoryESS{}, and oracle/RW centered
MAE is \BenchmarkWeakOracleCentered{}/\BenchmarkWeakRWCentered{}.  RW is not
better identified: raising exposures per user from 60 to 300 lowers oracle
error from \BenchmarkOracleAtSixty{} to \BenchmarkOracleAtThreeHundred{}.
All CE/RW benchmark variants borrow the known target grand mean, so their
absolute performance is an oracle-anchored shape diagnostic, not deployment
evidence.

\section{Discussion and Limitations}
\label{sec:discussion}

\noindent\textbf{What the audit establishes.}
Five distinctions are empirically consequential.  First, level discrepancy
is much larger than centered shape discrepancy.  Second, an impressive
absolute-MAE reduction can be produced by borrowing a random-insertion level anchor
while rank barely changes.  Third, exact video support does not eliminate
within-category video weighting; Kitagawa-style accounting makes that remaining
mix visible.  Fourth, restoring full support shows that excluded random support
is small on average but not zero.  Fifth, a frozen category calibration can
transport over one week while decision gains remain cutoff-specific.  Dashboard
audits should report these distinctions before recommending a correction or
acting on a ranking.

\noindent\textbf{What pairing does not establish.}
Shared user and video IDs do not make random-insertion and standard rows matched trials.
The logs can differ in context, position, time, user state, and exposure
frequency.  Random insertions support a reference for the retained realized
category--video mix; it does not randomize a video into another category.  The
within-video term in Eq.~\eqref{eq:decomp} is consequently a residual
accounting term, not a behavioral mechanism.  User-conditional exchangeability
for production adjustments remains untestable, and positivity is weak.  The
known-truth benchmark makes this boundary explicit: CE and RW can remove an
additive user level but cannot generally remove user--category interaction or
item selection; oracle weighting can in turn be variance-dominated under weak
overlap.

\noindent\textbf{Transport and measurement.}
The analysis covers one short-video platform and a 17-day matched window.
The $\geq5$ realized standard-coverage rule retains only
\EligibleStandardUserShare\% of base standard users, so estimates do not
transport automatically to less diversely exposed users.  Exact support also
discards roughly one quarter of random-insertion impressions and deliberately
omits support entry/exit from the primary gap.
Official KuaiRand categories are model predictions; the 0.8-confidence check
is reassuring for the level/shape conclusion but cannot eliminate label
error.  Category is fixed within video, so neither random insertions nor this
audit identifies pixel-level visual or advertising effects.  The prospective
test extends the calendar backward to April 8 but still uses one platform and
one adjacent test week.  Its returning paired-exposure cohort is selected by
post-period exposure presence, although not by outcomes, and retains only
\ProspectiveHistoryRetention\% of history-eligible users.  The semi-synthetic
benchmark is calibrated to category shares rather than a second platform; its
hand-set mechanism strengths are failure probes, not estimates of real serving
behavior.

\section{Conclusion}

On the same \PairedVideos{} videos, standard means differ from the random
reference mainly in level: absolute/centered MAE is
\NaiveAbs{}/\NaiveCentered{}, with rank \NaiveRank{}.  CE plus a random level
anchor lowers absolute MAE to \DemeanAbs{}, but shape and rank gains are modest.
Video mix receives \ItemMixShapleyShare\% of the absolute Shapley gap and
\CenteredItemMixShare\% of the centered gap; neither allocation is causal.
The frozen later test favors category over scalar calibration
(\ProspectiveCategoryAbs{} vs.\ \ProspectiveScalarAbs{} MAE), yet no
family-wise decision contrast excludes zero, including after disjoint
reference selection and valuation.  Known-truth tests show only partial
mean-additive improvement and failure under item mix or weak overlap.
Dashboard audits should match users and items, separate level, shape, and
decision loss, expose support and weighting, test transport, and keep claims
inside the logging design.

\FloatBarrier
\raggedbottom

\section{Ethical Considerations}

\noindent\textbf{Data and privacy.}\par\noindent
All real-data experiments use public, anonymized KuaiRand interactions; the
remaining experiments are synthetic.  We process no direct identifiers and
collect no new human-subject data.

\noindent\textbf{Potential misuse.}
Category rankings can concentrate exposure on engagement-maximizing content,
reducing supply diversity and disadvantaging long-tail creators.  Biased
rankings can also misallocate exposure while appearing objective.  We present
the output as a measurement audit, not a standalone allocation rule.
Deployment should combine uncertainty, diversity constraints, creator impact,
and objectives beyond engagement.

\bibliographystyle{ACM-Reference-Format}
\bibliography{refs}


\begin{thebibliography}{28}


\ifx \showCODEN    \undefined \def \showCODEN     #1{\unskip}     \fi
\ifx \showISBNx    \undefined \def \showISBNx     #1{\unskip}     \fi
\ifx \showISBNxiii \undefined \def \showISBNxiii  #1{\unskip}     \fi
\ifx \showISSN     \undefined \def \showISSN      #1{\unskip}     \fi
\ifx \showLCCN     \undefined \def \showLCCN      #1{\unskip}     \fi
\ifx \shownote     \undefined \def \shownote      #1{#1}          \fi
\ifx \showarticletitle \undefined \def \showarticletitle #1{#1}   \fi
\ifx \showURL      \undefined \def \showURL       {\relax}        \fi
\providecommand\bibfield[2]{#2}
\providecommand\bibinfo[2]{#2}
\providecommand\natexlab[1]{#1}
\providecommand\showeprint[2][]{arXiv:#2}

\bibitem[Andrews et~al\mbox{.}(2024)]%
        {andrews2024winners}
\bibfield{author}{\bibinfo{person}{Isaiah Andrews}, \bibinfo{person}{Toru
  Kitagawa}, {and} \bibinfo{person}{Adam McCloskey}.}
  \bibinfo{year}{2024}\natexlab{}.
\newblock \showarticletitle{Inference on Winners}.
\newblock \bibinfo{journal}{\emph{The Quarterly Journal of Economics}}
  \bibinfo{volume}{139}, \bibinfo{number}{1} (\bibinfo{year}{2024}),
  \bibinfo{pages}{305--358}.
\newblock
\href{https://doi.org/10.1093/qje/qjad043}{doi:\nolinkurl{10.1093/qje/qjad043}}


\bibitem[Ca{\~n}amares et~al\mbox{.}(2020)]%
        {canamares2020offline}
\bibfield{author}{\bibinfo{person}{Roc{\'{\i}}o Ca{\~n}amares},
  \bibinfo{person}{Pablo Castells}, {and} \bibinfo{person}{Alistair Moffat}.}
  \bibinfo{year}{2020}\natexlab{}.
\newblock \showarticletitle{Offline Evaluation Options for Recommender
  Systems}.
\newblock \bibinfo{journal}{\emph{Information Retrieval Journal}}
  \bibinfo{volume}{23}, \bibinfo{number}{4} (\bibinfo{year}{2020}),
  \bibinfo{pages}{387--410}.
\newblock
\href{https://doi.org/10.1007/s10791-020-09371-3}{doi:\nolinkurl{10.1007/s10791-020-09371-3}}


\bibitem[Castells and Moffat(2022)]%
        {castells2022offline}
\bibfield{author}{\bibinfo{person}{Pablo Castells} {and}
  \bibinfo{person}{Alistair Moffat}.} \bibinfo{year}{2022}\natexlab{}.
\newblock \showarticletitle{Offline Recommender System Evaluation: Challenges
  and New Directions}.
\newblock \bibinfo{journal}{\emph{AI Magazine}} \bibinfo{volume}{43},
  \bibinfo{number}{2} (\bibinfo{year}{2022}), \bibinfo{pages}{225--238}.
\newblock
\href{https://doi.org/10.1002/aaai.12051}{doi:\nolinkurl{10.1002/aaai.12051}}


\bibitem[Chaney et~al\mbox{.}(2018)]%
        {chaney2018algorithmic}
\bibfield{author}{\bibinfo{person}{Allison June-Barlow Chaney},
  \bibinfo{person}{Brandon~M. Stewart}, {and} \bibinfo{person}{Barbara~E.
  Engelhardt}.} \bibinfo{year}{2018}\natexlab{}.
\newblock \showarticletitle{How Algorithmic Confounding in Recommendation
  Systems Increases Homogeneity and Decreases Utility}. In
  \bibinfo{booktitle}{\emph{Proceedings of the 12th ACM Conference on
  Recommender Systems}}. \bibinfo{publisher}{ACM}, \bibinfo{pages}{224--232}.
\newblock
\href{https://doi.org/10.1145/3240323.3240370}{doi:\nolinkurl{10.1145/3240323.3240370}}


\bibitem[Crump et~al\mbox{.}(2009)]%
        {crump2009overlap}
\bibfield{author}{\bibinfo{person}{Richard~K. Crump},
  \bibinfo{person}{V.~Joseph Hotz}, \bibinfo{person}{Guido~W. Imbens}, {and}
  \bibinfo{person}{Oscar~A. Mitnik}.} \bibinfo{year}{2009}\natexlab{}.
\newblock \showarticletitle{Dealing with Limited Overlap in Estimation of
  Average Treatment Effects}.
\newblock \bibinfo{journal}{\emph{Biometrika}} \bibinfo{volume}{96},
  \bibinfo{number}{1} (\bibinfo{year}{2009}), \bibinfo{pages}{187--199}.
\newblock
\href{https://doi.org/10.1093/biomet/asn055}{doi:\nolinkurl{10.1093/biomet/asn055}}


\bibitem[Das~Gupta(1978)]%
        {dasgupta1978general}
\bibfield{author}{\bibinfo{person}{Prithwis Das~Gupta}.}
  \bibinfo{year}{1978}\natexlab{}.
\newblock \showarticletitle{A General Method of Decomposing a Difference
  Between Two Rates into Several Components}.
\newblock \bibinfo{journal}{\emph{Demography}} \bibinfo{volume}{15},
  \bibinfo{number}{1} (\bibinfo{year}{1978}), \bibinfo{pages}{99--112}.
\newblock
\href{https://doi.org/10.2307/2060493}{doi:\nolinkurl{10.2307/2060493}}


\bibitem[Deffayet et~al\mbox{.}(2023)]%
        {deffayet2023debiasedness}
\bibfield{author}{\bibinfo{person}{Romain Deffayet}, \bibinfo{person}{Philipp
  Hager}, \bibinfo{person}{Jean-Michel Renders}, {and} \bibinfo{person}{Maarten
  de Rijke}.} \bibinfo{year}{2023}\natexlab{}.
\newblock \showarticletitle{An Offline Metric for the Debiasedness of Click
  Models}. In \bibinfo{booktitle}{\emph{Proceedings of the 46th International
  ACM SIGIR Conference on Research and Development in Information Retrieval}}.
  \bibinfo{publisher}{ACM}, \bibinfo{pages}{558--568}.
\newblock
\href{https://doi.org/10.1145/3539618.3591639}{doi:\nolinkurl{10.1145/3539618.3591639}}


\bibitem[Dudoit et~al\mbox{.}(2004)]%
        {dudoit2004singlestep}
\bibfield{author}{\bibinfo{person}{Sandrine Dudoit}, \bibinfo{person}{Mark~J.
  van~der Laan}, {and} \bibinfo{person}{Katherine~S. Pollard}.}
  \bibinfo{year}{2004}\natexlab{}.
\newblock \showarticletitle{Multiple Testing. Part {I}. Single-Step Procedures
  for Control of General {Type I} Error Rates}.
\newblock \bibinfo{journal}{\emph{Statistical Applications in Genetics and
  Molecular Biology}} \bibinfo{volume}{3}, \bibinfo{number}{1}, Article
  \bibinfo{articleno}{13} (\bibinfo{year}{2004}), \bibinfo{numpages}{71}~pages.
\newblock
\href{https://doi.org/10.2202/1544-6115.1040}{doi:\nolinkurl{10.2202/1544-6115.1040}}


\bibitem[Gao et~al\mbox{.}(2022a)]%
        {gao2022kuairec}
\bibfield{author}{\bibinfo{person}{Chongming Gao}, \bibinfo{person}{Shijun Li},
  \bibinfo{person}{Wenqiang Lei}, \bibinfo{person}{Jiawei Chen},
  \bibinfo{person}{Biao Li}, \bibinfo{person}{Peng Jiang},
  \bibinfo{person}{Xiangnan He}, \bibinfo{person}{Jiaxin Mao}, {and}
  \bibinfo{person}{Tat-Seng Chua}.} \bibinfo{year}{2022}\natexlab{a}.
\newblock \showarticletitle{KuaiRec: A Fully-observed Dataset and Insights for
  Evaluating Recommender Systems}. In \bibinfo{booktitle}{\emph{Proceedings of
  the 31st ACM International Conference on Information \& Knowledge
  Management}}. \bibinfo{publisher}{ACM}, \bibinfo{pages}{540--550}.
\newblock
\href{https://doi.org/10.1145/3511808.3557220}{doi:\nolinkurl{10.1145/3511808.3557220}}


\bibitem[Gao et~al\mbox{.}(2022b)]%
        {gao2022kuairand}
\bibfield{author}{\bibinfo{person}{Chongming Gao}, \bibinfo{person}{Shijun Li},
  \bibinfo{person}{Yuan Zhang}, \bibinfo{person}{Jiawei Chen},
  \bibinfo{person}{Biao Li}, \bibinfo{person}{Wenqiang Lei},
  \bibinfo{person}{Peng Jiang}, {and} \bibinfo{person}{Xiangnan He}.}
  \bibinfo{year}{2022}\natexlab{b}.
\newblock \showarticletitle{KuaiRand: An Unbiased Sequential Recommendation
  Dataset with Randomly Exposed Videos}. In
  \bibinfo{booktitle}{\emph{Proceedings of the 31st ACM International
  Conference on Information \& Knowledge Management}}.
  \bibinfo{publisher}{ACM}, \bibinfo{pages}{3953--3957}.
\newblock
\href{https://doi.org/10.1145/3511808.3557624}{doi:\nolinkurl{10.1145/3511808.3557624}}


\bibitem[Gao et~al\mbox{.}(2024)]%
        {gao2024causalsurvey}
\bibfield{author}{\bibinfo{person}{Chen Gao}, \bibinfo{person}{Yu Zheng},
  \bibinfo{person}{Wenjie Wang}, \bibinfo{person}{Fuli Feng},
  \bibinfo{person}{Xiangnan He}, {and} \bibinfo{person}{Yong Li}.}
  \bibinfo{year}{2024}\natexlab{}.
\newblock \showarticletitle{Causal Inference in Recommender Systems: A Survey
  and Future Directions}.
\newblock \bibinfo{journal}{\emph{ACM Transactions on Information Systems}}
  \bibinfo{volume}{42}, \bibinfo{number}{4}, Article \bibinfo{articleno}{88}
  (\bibinfo{year}{2024}), \bibinfo{numpages}{32}~pages.
\newblock
\href{https://doi.org/10.1145/3639048}{doi:\nolinkurl{10.1145/3639048}}


\bibitem[Gilotte et~al\mbox{.}(2018)]%
        {gilotte2018offlineab}
\bibfield{author}{\bibinfo{person}{Alexandre Gilotte},
  \bibinfo{person}{Cl\'{e}ment Calauz\`{e}nes}, \bibinfo{person}{Thomas
  Nedelec}, \bibinfo{person}{Alexandre Abraham}, {and} \bibinfo{person}{Simon
  Doll\'{e}}.} \bibinfo{year}{2018}\natexlab{}.
\newblock \showarticletitle{Offline {A/B} Testing for Recommender Systems}. In
  \bibinfo{booktitle}{\emph{Proceedings of the Eleventh ACM International
  Conference on Web Search and Data Mining}}. \bibinfo{publisher}{ACM},
  \bibinfo{pages}{198--206}.
\newblock
\href{https://doi.org/10.1145/3159652.3159687}{doi:\nolinkurl{10.1145/3159652.3159687}}


\bibitem[Gruson et~al\mbox{.}(2019)]%
        {gruson2019offline}
\bibfield{author}{\bibinfo{person}{Alois Gruson}, \bibinfo{person}{Praveen
  Chandar}, \bibinfo{person}{Christophe Charbuillet}, \bibinfo{person}{James
  McInerney}, \bibinfo{person}{Samantha Hansen}, \bibinfo{person}{Damien
  Tardieu}, {and} \bibinfo{person}{Ben Carterette}.}
  \bibinfo{year}{2019}\natexlab{}.
\newblock \showarticletitle{Offline Evaluation to Make Decisions About Playlist
  Recommendation Algorithms}. In \bibinfo{booktitle}{\emph{Proceedings of the
  Twelfth ACM International Conference on Web Search and Data Mining}}.
  \bibinfo{publisher}{ACM}, \bibinfo{pages}{420--428}.
\newblock
\href{https://doi.org/10.1145/3289600.3291027}{doi:\nolinkurl{10.1145/3289600.3291027}}


\bibitem[Jeunen(2023)]%
        {jeunen2023position}
\bibfield{author}{\bibinfo{person}{Olivier Jeunen}.}
  \bibinfo{year}{2023}\natexlab{}.
\newblock \showarticletitle{A Probabilistic Position Bias Model for Short-Video
  Recommendation Feeds}. In \bibinfo{booktitle}{\emph{Proceedings of the 17th
  ACM Conference on Recommender Systems}}. \bibinfo{publisher}{ACM},
  \bibinfo{pages}{675--681}.
\newblock
\href{https://doi.org/10.1145/3604915.3608777}{doi:\nolinkurl{10.1145/3604915.3608777}}


\bibitem[Ji et~al\mbox{.}(2023)]%
        {ji2023dataleakage}
\bibfield{author}{\bibinfo{person}{Yitong Ji}, \bibinfo{person}{Aixin Sun},
  \bibinfo{person}{Jie Zhang}, {and} \bibinfo{person}{Chenliang Li}.}
  \bibinfo{year}{2023}\natexlab{}.
\newblock \showarticletitle{A Critical Study on Data Leakage in Recommender
  System Offline Evaluation}.
\newblock \bibinfo{journal}{\emph{ACM Transactions on Information Systems}}
  \bibinfo{volume}{41}, \bibinfo{number}{3}, Article \bibinfo{articleno}{75}
  (\bibinfo{year}{2023}), \bibinfo{numpages}{27}~pages.
\newblock
\href{https://doi.org/10.1145/3569930}{doi:\nolinkurl{10.1145/3569930}}


\bibitem[Joachims et~al\mbox{.}(2017)]%
        {joachims2017unbiasedltr}
\bibfield{author}{\bibinfo{person}{Thorsten Joachims}, \bibinfo{person}{Adith
  Swaminathan}, {and} \bibinfo{person}{Tobias Schnabel}.}
  \bibinfo{year}{2017}\natexlab{}.
\newblock \showarticletitle{Unbiased Learning-to-Rank with Biased Feedback}. In
  \bibinfo{booktitle}{\emph{Proceedings of the Tenth ACM International
  Conference on Web Search and Data Mining}}. \bibinfo{publisher}{ACM},
  \bibinfo{pages}{781--789}.
\newblock
\href{https://doi.org/10.1145/3018661.3018699}{doi:\nolinkurl{10.1145/3018661.3018699}}


\bibitem[Kitagawa(1955)]%
        {kitagawa1955components}
\bibfield{author}{\bibinfo{person}{Evelyn~M. Kitagawa}.}
  \bibinfo{year}{1955}\natexlab{}.
\newblock \showarticletitle{Components of a Difference Between Two Rates}.
\newblock \bibinfo{journal}{\emph{J. Amer. Statist. Assoc.}}
  \bibinfo{volume}{50}, \bibinfo{number}{272} (\bibinfo{year}{1955}),
  \bibinfo{pages}{1168--1194}.
\newblock
\href{https://doi.org/10.1080/01621459.1955.10501299}{doi:\nolinkurl{10.1080/01621459.1955.10501299}}


\bibitem[Li et~al\mbox{.}(2011)]%
        {li2011unbiasedoffline}
\bibfield{author}{\bibinfo{person}{Lihong Li}, \bibinfo{person}{Wei Chu},
  \bibinfo{person}{John Langford}, {and} \bibinfo{person}{Xuanhui Wang}.}
  \bibinfo{year}{2011}\natexlab{}.
\newblock \showarticletitle{Unbiased Offline Evaluation of
  Contextual-Bandit-Based News Article Recommendation Algorithms}. In
  \bibinfo{booktitle}{\emph{Proceedings of the Fourth ACM International
  Conference on Web Search and Data Mining}}. \bibinfo{publisher}{ACM},
  \bibinfo{pages}{297--306}.
\newblock
\href{https://doi.org/10.1145/1935826.1935878}{doi:\nolinkurl{10.1145/1935826.1935878}}


\bibitem[Parajuli et~al\mbox{.}(2026)]%
        {parajuli2026convergent}
\bibfield{author}{\bibinfo{person}{Sushobhan Parajuli},
  \bibinfo{person}{Samira~Vaez Barenji}, {and} \bibinfo{person}{Michael~D.
  Ekstrand}.} \bibinfo{year}{2026}\natexlab{}.
\newblock \showarticletitle{On the Convergent Validity of Offline Evaluation
  Designs for Recommender Systems}. In \bibinfo{booktitle}{\emph{Proceedings of
  the 20th ACM Conference on Recommender Systems}}. \bibinfo{publisher}{ACM}.
\newblock
\href{https://doi.org/10.1145/3773078.3831818}{doi:\nolinkurl{10.1145/3773078.3831818}}
\newblock
\shownote{Forthcoming}.


\bibitem[Sachdeva et~al\mbox{.}(2020)]%
        {sachdeva2020deficientsupport}
\bibfield{author}{\bibinfo{person}{Noveen Sachdeva}, \bibinfo{person}{Yi Su},
  {and} \bibinfo{person}{Thorsten Joachims}.} \bibinfo{year}{2020}\natexlab{}.
\newblock \showarticletitle{Off-Policy Bandits with Deficient Support}. In
  \bibinfo{booktitle}{\emph{Proceedings of the 26th ACM SIGKDD International
  Conference on Knowledge Discovery \& Data Mining}}. \bibinfo{publisher}{ACM},
  \bibinfo{pages}{965--975}.
\newblock
\href{https://doi.org/10.1145/3394486.3403139}{doi:\nolinkurl{10.1145/3394486.3403139}}


\bibitem[Saito et~al\mbox{.}(2021)]%
        {saito2021openbandit}
\bibfield{author}{\bibinfo{person}{Yuta Saito}, \bibinfo{person}{Shunsuke
  Aihara}, \bibinfo{person}{Megumi Matsutani}, {and} \bibinfo{person}{Yusuke
  Narita}.} \bibinfo{year}{2021}\natexlab{}.
\newblock \showarticletitle{Open Bandit Dataset and Pipeline: Towards Realistic
  and Reproducible Off-Policy Evaluation}. In
  \bibinfo{booktitle}{\emph{Proceedings of the Neural Information Processing
  Systems Track on Datasets and Benchmarks}},
  \bibfield{editor}{\bibinfo{person}{Joaquin Vanschoren} {and}
  \bibinfo{person}{Sai-Kit Yeung}} (Eds.), Vol.~\bibinfo{volume}{1}.
\newblock
\urldef\tempurl%
\url{https://datasets-benchmarks-proceedings.neurips.cc/paper/2021/hash/33e75ff09dd601bbe69f351039152189-Abstract-round2.html}
\showURL{%
\tempurl}


\bibitem[Saito and Joachims(2022)]%
        {saito2022largeactionope}
\bibfield{author}{\bibinfo{person}{Yuta Saito} {and} \bibinfo{person}{Thorsten
  Joachims}.} \bibinfo{year}{2022}\natexlab{}.
\newblock \showarticletitle{Off-Policy Evaluation for Large Action Spaces via
  Embeddings}. In \bibinfo{booktitle}{\emph{Proceedings of the 39th
  International Conference on Machine Learning}}
  \emph{(\bibinfo{series}{Proceedings of Machine Learning Research},
  Vol.~\bibinfo{volume}{162})}. \bibinfo{publisher}{PMLR},
  \bibinfo{pages}{19089--19122}.
\newblock
\urldef\tempurl%
\url{https://proceedings.mlr.press/v162/saito22a.html}
\showURL{%
\tempurl}


\bibitem[Saito et~al\mbox{.}(2020)]%
        {saito2020unbiased}
\bibfield{author}{\bibinfo{person}{Yuta Saito}, \bibinfo{person}{Suguru
  Yaginuma}, \bibinfo{person}{Yuta Nishino}, \bibinfo{person}{Hayato Sakata},
  {and} \bibinfo{person}{Kazuhide Nakata}.} \bibinfo{year}{2020}\natexlab{}.
\newblock \showarticletitle{Unbiased Recommender Learning from
  Missing-Not-At-Random Implicit Feedback}. In
  \bibinfo{booktitle}{\emph{Proceedings of the 13th International Conference on
  Web Search and Data Mining}}. \bibinfo{publisher}{ACM},
  \bibinfo{pages}{501--509}.
\newblock
\href{https://doi.org/10.1145/3336191.3371783}{doi:\nolinkurl{10.1145/3336191.3371783}}


\bibitem[Schnabel et~al\mbox{.}(2016)]%
        {schnabel2016recommendations}
\bibfield{author}{\bibinfo{person}{Tobias Schnabel}, \bibinfo{person}{Adith
  Swaminathan}, \bibinfo{person}{Ashudeep Singh}, \bibinfo{person}{Navin
  Chandak}, {and} \bibinfo{person}{Thorsten Joachims}.}
  \bibinfo{year}{2016}\natexlab{}.
\newblock \showarticletitle{Recommendations as Treatments: Debiasing Learning
  and Evaluation}. In \bibinfo{booktitle}{\emph{Proceedings of the 33rd
  International Conference on Machine Learning}}
  \emph{(\bibinfo{series}{Proceedings of Machine Learning Research},
  Vol.~\bibinfo{volume}{48})}. \bibinfo{publisher}{PMLR},
  \bibinfo{pages}{1670--1679}.
\newblock
\urldef\tempurl%
\url{https://proceedings.mlr.press/v48/schnabel16.html}
\showURL{%
\tempurl}


\bibitem[Shorrocks(2013)]%
        {shorrocks2013decomposition}
\bibfield{author}{\bibinfo{person}{Anthony~F. Shorrocks}.}
  \bibinfo{year}{2013}\natexlab{}.
\newblock \showarticletitle{Decomposition Procedures for Distributional
  Analysis: A Unified Framework Based on the {Shapley} Value}.
\newblock \bibinfo{journal}{\emph{The Journal of Economic Inequality}}
  \bibinfo{volume}{11}, \bibinfo{number}{1} (\bibinfo{year}{2013}),
  \bibinfo{pages}{99--126}.
\newblock
\href{https://doi.org/10.1007/s10888-011-9214-z}{doi:\nolinkurl{10.1007/s10888-011-9214-z}}


\bibitem[Vrijenhoek et~al\mbox{.}(2025)]%
        {vrijenhoek2024radio}
\bibfield{author}{\bibinfo{person}{Sanne Vrijenhoek}, \bibinfo{person}{Gabriel
  B\'{e}n\'{e}dict}, \bibinfo{person}{Mateo Gutierrez~Granada}, {and}
  \bibinfo{person}{Daan Odijk}.} \bibinfo{year}{2025}\natexlab{}.
\newblock \showarticletitle{{RADio*} -- An Introduction to Measuring Normative
  Diversity in News Recommendations}.
\newblock \bibinfo{journal}{\emph{ACM Transactions on Recommender Systems}}
  \bibinfo{volume}{3}, \bibinfo{number}{1}, Article \bibinfo{articleno}{5}
  (\bibinfo{year}{2025}), \bibinfo{numpages}{29}~pages.
\newblock
\href{https://doi.org/10.1145/3636465}{doi:\nolinkurl{10.1145/3636465}}


\bibitem[Wang et~al\mbox{.}(2019)]%
        {wang2019doublyrobust}
\bibfield{author}{\bibinfo{person}{Xiaojie Wang}, \bibinfo{person}{Rui Zhang},
  \bibinfo{person}{Yu Sun}, {and} \bibinfo{person}{Jianzhong Qi}.}
  \bibinfo{year}{2019}\natexlab{}.
\newblock \showarticletitle{Doubly Robust Joint Learning for Recommendation on
  Data Missing Not at Random}. In \bibinfo{booktitle}{\emph{Proceedings of the
  36th International Conference on Machine Learning}}
  \emph{(\bibinfo{series}{Proceedings of Machine Learning Research},
  Vol.~\bibinfo{volume}{97})}. \bibinfo{publisher}{PMLR},
  \bibinfo{address}{Long Beach, California, USA}, \bibinfo{pages}{6638--6647}.
\newblock
\urldef\tempurl%
\url{https://proceedings.mlr.press/v97/wang19n.html}
\showURL{%
\tempurl}


\bibitem[Yang et~al\mbox{.}(2018)]%
        {yang2018unbiased}
\bibfield{author}{\bibinfo{person}{Longqi Yang}, \bibinfo{person}{Yin Cui},
  \bibinfo{person}{Yuan Xuan}, \bibinfo{person}{Chenyang Wang},
  \bibinfo{person}{Serge Belongie}, {and} \bibinfo{person}{Deborah Estrin}.}
  \bibinfo{year}{2018}\natexlab{}.
\newblock \showarticletitle{Unbiased Offline Recommender Evaluation for
  Missing-Not-At-Random Implicit Feedback}. In
  \bibinfo{booktitle}{\emph{Proceedings of the 12th ACM Conference on
  Recommender Systems}}. \bibinfo{publisher}{ACM}, \bibinfo{pages}{279--287}.
\newblock
\href{https://doi.org/10.1145/3240323.3240355}{doi:\nolinkurl{10.1145/3240323.3240355}}


\end{thebibliography}

\end{document}